\documentclass[10pt,conference]{IEEEtran}
\usepackage[dvipsnames]{xcolor}
\usepackage[english]{babel}
\usepackage{colortbl}
\usepackage{blindtext}
\usepackage{tikz}
\usepackage{amsmath}
\usepackage{tabu}
\usepackage{adjustbox}
\usepackage{xspace}
\usepackage{multirow}
\usepackage{multicol}
\usepackage[font=small]{caption} % decreasing the font size of the caption of table or figure
\usepackage{float}
\usepackage{csquotes}
\usepackage{textcomp}

\usepackage{lipsum}
\usepackage{mwe}
\usepackage{url}
\usepackage[boxed,vlined]{algorithm2e}
\usepackage{paralist} % also compact lists
\usepackage{listings} % to get sql etc nicely formatted
\usepackage{bbding} % for the \CheckmarkBold symbol
\usepackage[draft]{changes} % allow for orange-suggestions
\usepackage{hyperref}
\usepackage{array}
\newcolumntype{P}[1]{>{\centering\arraybackslash}p{#1}}

\usepackage{diagbox}
\DeclareUnicodeCharacter{2212}{-}

\newcommand{\hide}[1]{}

\newcommand{\bit}{\begin{compactitem}}
\newcommand{\eit}{\end{compactitem}}
\newcommand{\ben}{\begin{compactenum}}
\newcommand{\een}{\end{compactenum}}

\newcommand{\method}{Repo2Vec\xspace}
\newcommand{\meta}{Repo2Vec\_M\xspace}   % Metadata
\newcommand{\metanode}{Repo2Vec\_MS\xspace} %Metadata and structure
\newcommand{\metanodecode}{Repo2Vec\_All\xspace}  % All three
\newcommand{\github}{GitHub\xspace}
\newcommand{\TotalRepoNum}{1013\xspace}
\newcommand{\dben}{D\_ben\xspace}
\newcommand{\dmal}{D\_mal\xspace}

\newcommand{\ofr}[1]{{\textcolor{blue}{\bf OFR:}}
{\textcolor{purple}{\bf #1}}}

\newcommand{\miii}[1]{{\textcolor{purple}{\bf MF:}}
{\textcolor{blue}{\bf #1}}}

\newcommand{\risul}[1]{{\textcolor{green}{\bf Risul:}}
{\textcolor{orange}{\bf #1}}}

\newcommand{\eatreminders}{
\renewcommand{\ofr}[1]{}
\renewcommand{\miii}[1]{}
\renewcommand{\risul}[1]{}
}
\eatreminders

\begin{document}

\title{\method: A Comprehensive Embedding Approach for Determining Repository Similarity}
% Similar Software Repositories in GitHub 

\author{
{\rm Md Omar Faruk Rokon}\\
UC Riverside\\
mroko001@ucr.edu
% copy the following lines to add more authors
\and
{\rm Pei Yan}\\
UC Riverside\\
pyan012@ucr.edu
\and
{\rm Risul Islam}\\
UC Riverside\\
risla002@ucr.edu
\and 
{\rm Michalis Faloutsos}\\ 
UC Riverside\\
michalis@cs.ucr.edu
} % 

\maketitle

\thispagestyle{plain}
\pagestyle{plain}

%\section{Acknowledgments}
%We would like to thank blah blah

%\begin{abstract}

%%%%%%%%%%%%%%%%%%%%%%%%%%%%%%%%%%%%%%%
% AUTHOR: Christos Faloutsos
% INSTITUTION: CMU
% DATE: April 2019
% GOAL: to streamline the paper presentations
%%%%%%%%%%%%%%%%%%%%%%%%%%%%%%%%%%%%%%%

% \comment{test}
%\notice{{\em rhetorical question:} - What is the best rhetorical question you can start with?}
\begin{abstract}
How can we identify similar repositories and clusters among a large online archive, such as GitHub? 
% Many online archives, such as GitHub, encourage  developers to  copy, reuse, and improve the existing projects.
Determining repository similarity is an essential building block in studying the dynamics and the evolution of such software ecosystems. 
%  task for developers where Machine Learning (ML) approaches can play a vital role. 
The key challenge is to determine the right representation for the diverse repository  features in a way that: (a) it captures all aspects of  the available  information, and (b) it is readily usable by ML algorithms. 
%numeric features maintaining the semantics of the content of a repository. %  However, to use standard ML approaches, we need to represent a repository in numeric features maintaining the semantics of the content of a repository. 
% %\notice{{\em 'what' - NOT 'how':} 
%List the benefits of the approach - NOT the details of how you do it!}
We propose \method, a comprehensive embedding approach to represent a repository as a distributed vector by combining features from three  types of information sources.
%: (a) metadata information, (b) source code, and (c) software directory structure.
As our key novelty, we consider three types of information: (a) metadata, (b) 
the structure of the repository, and (c) the source code.
We also introduce a series of embedding approaches
to represent and combine these information types
into a single embedding.
% \notice{{\em baptize:} Give a NAME to the method - ideal name should
% (a) be an english-like word, but NOT a vocabulary word 
% (b) easy to pronounce ('say it three times, quickly')
% (c) should emphasize the main idea/insight/advantage of your method 
%     (NOT the steps you took - 'FraudSpot' is good, 'DeepLearnFraud' is not) 
% (d) should have positive connotation 
%     ('eagle', 'lion', 'safe', 'guard', 'spot', 'alert')
% } 
% The challenge is three fold: (a) identifying the right features from each type of information, (b) combining the 
% By considering all these types of information, we argue that our
% representation captures the semantic and structural  meaning of a  repository comprehensively.
% , and can be used as features in finding similarity, and clusters, and classification task.  
% (a) {\em \scale}, being linear on the input size
% (b) {\em \effective}, spotting 90\% of the anomalies in real data
% (c) {\em \automatic}, requiring no user-defined parameters.
%\notice{{\em numbers}: Mention some performance numbers}
We evaluate our method with two real datasets from \github for a combined \TotalRepoNum repositories.  
%using a combination of ground-truth and manual inspection.
First, we show that our method outperforms previous methods in terms of precision (93\% vs 78\%),
with  nearly  twice  as  many  Strongly  Similar  repositories  and  30\%  fewer False Positives.
%for the same rate of query success.
Second, we show how \method provides a solid basis for: 
(a) distinguishing between malware and benign repositories, and (b) identifying a meaningful hierarchical clustering. 
For example, we achieve 98\% precision, and 96\% recall in distinguishing malware and benign repositories.  
% Third,  our method can form the basis for a hierarchical clustering 
% that captures accurately the  purpose of the repositories in each cluster.
% \miii{refined statement}
Overall, our work is a fundamental building block for enabling many repository analysis functions such as repository categorization by target platform or intention, detecting code-reuse and clones, and identifying lineage and evolution.
\end{abstract}
%, and clone detection

% Experiments on two dataset of java repositories
% illustrate the benefits of our method. First, \method can identify similarity and dissimilarity among repositories with 100\% success rate and 93\% precision which outperforms previous methods. Second, applying the method in a malware repositories, we find a hierarchical clusters of malware repositories that aligns with their purpose and lineage. Finally, the embedding vector that we offer is ready to be used for various downstream task e.g. repository category classification, malware classification, tagging/captioning, and clone detection.   

% \reminder{
% Revision: \svnrev ;
% On \svndate
% }
%\end{abstract}

\section{Introduction}
\label{sec:intro}

%{\bf Context:}
%MF: 
Establishing a way to measure similarity between software repositories is an essential building block for studying the plethora of  repositories in online Open Source Software (OSS) platforms. 
These OSS platforms contain a massive number of repositories and engagement of millions of users~\cite{spinellis2020commit}.
There are significant collaborations and code reuses~\cite{spinellis2020dataset, gharehyazie2017some} on these platforms, which are openly supported and encouraged.
Researchers are interested in studying the dynamics of such repositories, which include the ability to
identify: (a) derivative repositories, (b) families of repositories, (c) the evolution of software projects, and (d) coding and technology trends.
 \github is arguably the largest such platform with more than 32 million repositories and 34 million users exhibiting significant collaborative interactions~\cite{mockus2007large}. 

\begin{figure}[t]
    \begin{center}
        \begin{tabular}{c}
           \includegraphics[width=0.45\textwidth, height=5.5cm]{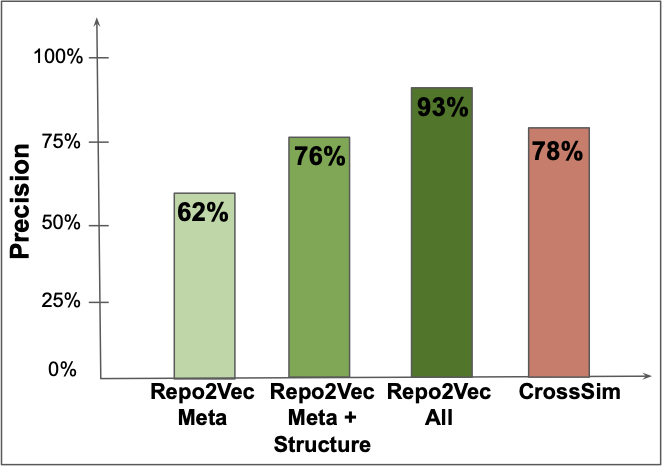} \\
        \end{tabular}
        \caption{Our approach outperforms the state of the art approach CrossSim in terms of precision using CrossSim dataset. We also see the effect of different types of information that \method considers:  metadata, adding structure, and adding source code information. 
           \label{fig:crown_jeweel} }
    \end{center}
\end{figure}

% -----
% More than 50\% of source code files of open source projects are copied from other projects~\cite{mockus2007large} which implies that a large part of GitHub repositories are similar. GitHub is the largest open source software repository system containing more than 32 million repositories and 34 million users. It hosts various kinds of open source software projects which gives developers an ample opportunity to observe, learn, and reuse the source code projects.  While code reuse is an opportunity to improve the software quality and lessen the workload to create complex software, it also alarming for code security as it might include loopholes in the software.  

% GitHub offers a search option to search repositories, issues, and user name related to a query. It matches text present in the repositories and retrieves the relevant repositories. As providing  text in a repository is optional for developer, it is not ideal to find similar repositories.  The text might also be irrelevant to project as GitHub does not impose any restrictions. Moreover, it does not care the actual contents of the repositories like source files and project structure. 

% -----

%{\bf Problem:} 
How can we quantify the level of similarity between two repositories?
This is the problem that we address here.
Focusing on GitHub, every repository consists of  metadata,  source code, and auxiliary files. Given a repository, how can we  identify the most similar repositories among a large set? 
The input here is a large number of repositories and a set of queries. The desired output is: (a) the most similar repositories for a given query repository and (b) clusters of similar repositories. The key challenge here is to represent the repository data into a numeric feature vectors to enable ML approaches to compute the similarities and cluster among repositories. In addition, combining vectors from different types of information, as we will do here, is also a challenge. 

%{\bf Prior work:}
There are relatively few efforts that focus on establishing similarity between repositories, and most of them use either  metadata or source code level information, while none of them
use the three types of information that we do here. 
First,  LibRec~\cite{thung2013automated}, SimApp~\cite{chen2015simapp}, Collaborative Tagging~\cite{thung2012detecting}, and RepoPal~\cite{zhang2017detecting} utilize only metadata to find similarity among repositories.
Second, MUDABLUE~\cite{kawaguchi2006mudablue} and CLAN~\cite{mcmillan2012detecting} are two similarity computation approaches using only source code of repositories as plain text. 
Third, CrossSim~\cite{nguyen2018crosssim, nguyen2020automated} proposes a graph representation to compute similarity between repositories using both metadata and source code. 
We discuss the related works in more detail in Section~\ref{sec:related_works}.

%MOVE OUT: Although the prior works perform well for a set of certain contexts, there are some limitation and challenges. None has taken advantages of project directory structure to compute similarity between two repositories. No approaches has preserved and employed the semantic meaning of source code and metadata. Also, they do not offer a standard model which will enable ML approach to work on GitHub repositories. 

%{\bf Contribution:}
As our key contribution,
we propose \method, an embedding approach to represent software repositories with a multi-dimensional distributed continuous vector which can be used to measure the similarity between repositories. 
%Our approach is novel for two reasons. 
We briefly describe the key features of our approach.
First,
our method represents a repository as a distributed continuous vector in an embedding space.
%\miii{Didn't people complain about this us being the first? I bet some previous methods use some embedding, no? If not, you should explain: I think though it becomes a tricky point: code2vec uses embedding...}
Second,
we consider three types of information:  (a) metadata, (b) source code, and (c) the repository directory structure. 
% it is tricky but I think this is good...  still confusing to me  -- fair enough, what do you propose? repository directory structure alright, go with yours, I am moving on to next section. go to evaluation 
%One other thing: having repo2vec all small, makes it ugly to start a sentence with it. Why not capitcalize Repo2Vec?  no problem  :-) Off to evaluation, but lunch may interject, I will be here Take your time. Risul did a great job at evaluation
% \miii{Can we say that using embedding is new?}yes FOR REAL? yes for real Ok, let's "risk it" and go all out.
Our approach provides a flexible way to combine these three types
 using our default values which can be customized to match
the niche needs of a savvy user.
The significance of our approach is that it 
generates a relatively-low dimensional vector 
that can enable follow up repository analysis.
Such follow up studies can leverage the plethora of ML techniques:
we provide a proof of concept for two such applications here.

\miii{Also do we want to mention here or in Discussion: that our approach can also enable a savvy user to give different "weights" on the different sources of information depending on application?}

We deploy our approach and study the similarity on a malware dataset of 433 repositories and a benign dataset of 580 repositories. First, we demonstrate the effectiveness of our method by comparing  it 
% similarity performance 
against state of the art works. 
Second, we show how our \method can enable algorithms for: 
(a) distinguishing between malware and benign repositories, and (b) identifying a meaningful hierarchical clustering.  The key results are briefly discussed below.

%classification dataset is differentye
{\bf a. \method outperforms prior works.} 
For this comparison, we select  the best approach to date, CrossSim, which has been shown to outperform previous approaches~\cite{zhang2017detecting, kawaguchi2006mudablue, mcmillan2012detecting}.
For consistency, we also follow their evaluation methodology and use their
dataset with 580 benign repository.
We show that our approach identifies similar repositories  with 93\% precision compared to 78\%  as shown in Figure~\ref{fig:crown_jeweel}. 
Further, our approach finds nearly {\bf twice as many strongly similar} repositories  and 30\% fewer False Positives, as we see in Figure~\ref{fig:confidence_compare}.

% our  method  outperforms  previous  methods  in  termsof  precision  (93\%  vs  78\%)  for  the  same  rate  of  query  success, which is 100\% for both methods.
%We even get higher precision 100\% for malware repository dataset.

{\bf b.   Metadata and structure  provide significant performance.} We assess the information contribution of three types of information. Interestingly, we can identify similarity fairly well without the use of source code   as shown in Figure~\ref{fig:crown_jeweel}. 
Using only metadata and structure leads to a 76\% precision, which is comparable to the previous best method, which uses source code. 

% ---
% c. (Observations from large scale application): We find that there is a skewed distribution of "similarity repos clusters"...  Applying it on X repos, we find 100 clusters of highly skewed sizes: largest cluster is keyloggers, and ransomware.
% ---
{\bf c. Application: identifying malware repositories accurately.}
We show that our approach can enable a supervised classification approach.
We focus on distinguishing malware from benign repositories, which is
a practical problem~\cite{rokon2020sourcefinder}.
Using our embedding,
we can identify malware repositories with 98\% precision and 97\% recall, which outperforms the previous approaches.

{\bf d. Application: identifying a meaningful hierarchy.} 
We show that our approach can form the basis for 
a meaningful (unsupervised)
hierarchical clustering of repositories. We show that
the emerging  structure aligns with their purpose and lineage.
In our evaluation, we  focus at two levels of granularity: a coarse and a fine level with 3 and 26 clusters respectively.
Using an LDA-based topic extraction method,
we find that the clusters are cohesive:
more than 80\%  of the repositories per cluster have the same focus. 
We discuss the clustering in Section \ref{sec:application}.

{\em Our work in perspective.} 
Our approach can be seen as
a first step towards the use of embedding approaches in repository analysis. In fact, it can be seen as a general framework where the selection of individual features can be driven by the intention of the application. For example,
one can focus on different primary features depending on whether we want to identify: (a) plagiarism or function level similarity, 
(b) programming styles, or (c) software intention.
We intend to share our method and our dataset to facilitate follow up research in this direction.

\section{Background}
\label{sec:background}

%%%%%%%%%%%%%%%%%%%%%%%%%%%%%%%%%%%%%%%
% AUTHOR: Christos Faloutsos
% INSTITUTION: CMU
% DATE: April 2019
% GOAL: to streamline the paper presentations
%%%%%%%%%%%%%%%%%%%%%%%%%%%%%%%%%%%%%%%
We provide  some background on \github and describe embedding approaches, which we extend and use later.
% such as word2vec, doc2vec, code2vec, and node2vec in this Section.

{\bf A. \github and its features.}
GitHub is a massive software archive, which
enables  users to store and share code
creating a global social network of interaction. Users can collaborate on a repository by raising issues or forking projects, where they copy and evolve projects. Users can follow projects, and ``up-vote" projects using ``stars". 
%Although \github has many private repositories, there are 32 million public software repositories in there.
We describe the key elements of a \github repository here. A repository contains three types of information (a) metadata, (b) project directory, and (c) source code files, which we explain below. 

{\bf a. Metadata}: A repository in GitHub has a large number of metadata fields. Most notable are: (a) title, (b) descriptions, (c) topics, and (d) readme file. All these fields are optional and they are provided by 
the author. Commit and issues are other sources of textual metadata which include messages about the specific functionality of the repository. 
At the same time, there are  metrics that capture the popularity of a repository including:  (a) stars, (b) forks, and (c) watches. As the text fields are provided by the repository author, they can be unstructured,
noisy, or missing altogether.

{\bf b. Source code}: It is the core element of a software repository. A repository contains software projects written in various programming languages such as C/C++, Java, Python, and so on. These source codes are the logical centre of a software stored in a repository. 

{\bf c. Project directory structure}: A well-crafted software repository follows a best-practices directory structure containing dataset, source code, and other auxiliary files. We hypothesize that the structure could be useful in establishing similarity between repositories. 
%\miii{This phrase is awkward: is it a fact, a hypothesis, or we prove that later? it is fact -- who facted it out? :-) paper?no - so it is fact in your mind! :-) Should we leave it as I wrote it? If not, revise. its fine. may be we can change the word "hypothesize"} 

%\miii{I wonder if we could explain some of the features a bit more, as now this conference is not focused on repos...Just a thought.}

{\bf B. Embedding approaches.}
An embedding (a.k.a. distributed representation) is 
an unsupervised approach for mapping entities, such as words or images,  into a relatively low-dimensional space by using a deep neural network on a large training corpus~\cite{le2014distributed, mikolov2013distributed}. 
Although the method is unsupervised, it relies on ideally a large dataset,
which is used to ``train" the neural network. The neural network develops a model of the dataset, which we can think of as probabilities and correlations of its entities.
Embedding approaches have revolutionized research in several fields, such as Natural Language Processing (NLP)~\cite{le2014distributed, mikolov2013distributed,pennington2014glove,peters2018deep}, computer vision~\cite{kottur2016visual}, graph mining~\cite{narayanan2017graph2vec, grover2016node2vec}, and software analysis~\cite{alon2019code2vec}.  % Let's 

The power of an embedding is twofold:
(a) it can simplify the representation of a complex entity with diverse features, including categorical, and (b) it provides a way to quantify entity similarity as a function of the distance of their corresponding vectors. 
%what domains do these refer to: add the domains, image and video processing?
An efficient embedding has the following properties: (a) it gives a fixed and low dimensional vector, and (b) it ensures that semantically similar objects are  ``close by" in the embedding space.

{\bf a. Word embedding: word2vec.}
In the seminal word2vec work~\cite{mikolov2013distributed},
we map words to vectors in a way that similar words, such as ``father'' and ``parent'', map to nearby vectors.
This similarity is established by ``feeding" a large corpus of documents to the deep
neural network.
In other words, the model captures word correlations by
calculating the probability with which a 
word can appear within a given neighborhood of words.

{\bf b. Document embedding: doc2vec.}
The doc2vec~\cite{le2014distributed} is an unsupervised embedding model for a variable length paragraph or document.
%\risul{Can we be more clear or check}
The model takes a document as input and maps it to an M-dimensional embedding vectors while doing a proxy task, predicting target word or sampled words in the document.
%\miii{This part I don't understand: does it map it or does it predict?}
%by concatenating the document vector with several word vectors. 
%This model is based on the word embedding~\cite{mikolov2013distributed} model. 

% \begin{figure}
%     \begin{center}
%         \begin{tabular}{c}
%           \includegraphics[width=0.45\textwidth]{FIG/docvec_architecture.png} \\
%         \end{tabular}
%         \caption{Overview of the doc2vec architecture.
%           \label{fig:doc2vec} }
%     \end{center}
% \end{figure}

%{\bf The Architecture of doc2vec: }
In more detail, the document  %or paragraph paragraph is a document, no?
embedding model is based on the word embedding~\cite{mikolov2013distributed} model. The main difference between them is the introduction of  the document id vector.
%\miii{Is this a learned vector or a randomly selected id?}.  
Like word2vec, there are two types of doc2vec available:
%\miii{Two types or word2vec can be implemented in two different ways? ANd is this Important here? Did you mentiond int the word2vec?}: 
(a) Distributed Memory Model of Paragraph Vectors (PV-DM) and (b) Distributed Bag of Words version of Paragraph Vector (PV-DBOW). 
PV-DM is similar to the Continuous Bag of Words (CBOW) model in word2vec. 
 %with the addition of document id.
%, and document vector is the product of this task. 
The PV-DBOW model is similar to the skip-gram model of word2vec. %which is trained to predict sampled words from a document. 
The document vector is calculated at the same time as the word vectors of the document.
% \miii{Needs some explanation... or leave as is}
Note that, PV-DM performs better for large, and well-structured documents. On the other hand, PV-DBOW is considered a better choice for small and defective documents, as it is computationally fast. 

%\miii{Somewhat confusing: again the predicting a word confuses me...}
%The architecture consists of document matrices {\em (D)} and word matrices {\em (W)} where a document is mapped to a column in {\em D}, and a word is mapped to a column in {\em W}. These matrices are averaged or concatenated to predict next word in the context. For example, concatenation of the word matrices for ``the", ``model", ``can", and the document matrix is used to predict the fourth word ``be". Figure \ref{fig:doc2vec} shows the architecture of doc2vec model. We refer to the original paper~\cite{le2014distributed} for more details about doc2vec.

{\bf c. Code embedding: code2vec.}
Embedding approaches have also been proposed for detecting code similarity.
A recent approach is code2vec
which maps a method (or more generally a code snippet) of  arbitrary length 
%\miii{Why snippet, we can say source code, they actually did this on a "function" }
%The code2vec~\cite{alon2019code2vec} is a neural attention-based embedding approach\miii{The way you describe it sounds different than the others before. Is it? If yes, how?}. 
to an M-dimensional vector~\cite{alon2019code2vec,compton2020embedding}. 
The code2vec approach uses program structure explicitly
to predicting program properties and uses an attention based neural network that learns 
a  continuous
distributed vector representation for the code snippet.
As a result, one can compare and group code snippets.
The process is fairly involved as it attempts to capture
the logical structure and flow of the program and the sequence of commands.
For example, the code is decomposed into a set of paths based on its
abstract syntax tree. The neural network learns simultaneously:
the representation of each path and how to aggregate a set of them.
Due to space limitations, we refer the interested reader to the original work~\cite{alon2019code2vec}.

{\bf d. Node embedding: node2vec.}
%\miii{I want to stress it is node mapping so that our transformation of struct2vec is more than node2vec!}
The node2vec~\cite{grover2016node2vec} is a graph embedding approach for mapping a node in a network to an M-dimensional embedding vector. The model  maximizes the likelihood of preserving network neighborhoods of nodes using Stochastic Gradient Descent (SGD).

% \begin{figure}
%     \begin{center}
%         \begin{tabular}{c}
%           \includegraphics[width=0.45\textwidth]{FIG/nodevec_architecture.png} \\
%         \end{tabular}
%          \caption{Overview of the node2vec architecture: (a)  creates node sequences from an input Graph using biased random walk, (b) trains with skip-gram model, (c) computes the embedding for a node. \label{fig:node2vec}}
%     \end{center}
% \end{figure}

%{\bf The architecture of node2vec:}
In more detail, the model computes the embedding based on nodes neighborhoods. First, the network structure is converted to a set of paths (node sequences) using a biased random walk sampling strategy which combines Depth-First Sampling (DFS) and Breadth-First Sampling (BFS) for every nodes. The sampling strategy efficiently explores diverse neighborhoods of a given node.
%combines with Depth-first Sampling (DFS) and (Breadth-first Sampling) BFS \miii{spell out} search algorithms. 
%Many different neighborhoods are sampled for each node, neighborhoods are not restricted to just immediate neighbors but can have large structures. 
These sets of paths can be analogized to the sentences in a document. 
%for traditional text processing. 
Then the model is trained on these node sequences with the skip-gram models presented in word2vec~\cite{mikolov2013distributed} to get the vector representations for each node.  %Later node2vec computes embedding for each node in the graph. 
%The node2vec model is depicted in Figure \ref{fig:node2vec}.
For more details about the model, we refer to the original paper~\cite{grover2016node2vec}.
%\miii{The description needs work: structure is not converted into node sequences: we capture the network structure with a set of paths. These paths are a subset of all possible walks on the graph using a biased random walk sampling strategy which combines...Do you do these for every node in the graph? Say so. How many paths per node? }

%\ofr{Next two sentences are confusing. How they train them using skip-gram? How node2vec computes embedding?} 

% However none of the above methods fullfils all the specs
% of our method: (a) scalability (b) effectiveness.
% \notice{{\em salesman matrix}: last *column*: YOUR method}
% Table~\ref{tab:salesman} contrasts \method against
% the state of the art competitors.

% \begin{table}[htbp]
% \begin{center}
% \begin{tabular}{ l| c | c||c|}
%       \diagbox{Property}{Method}    & \rotatebox{80}{method1} & \rotatebox{80}{method2} & \rotatebox{80}{\method} \\ 
% \hline  
% 	\scale &  &  & \CheckmarkBold \\ 
%         \effective    &  &  & \CheckmarkBold \\ 
%         \automatic &  &  & \CheckmarkBold \\ 
%         other-stuff &  &  & \CheckmarkBold \\ \hline
% \end{tabular} 
% \caption{ {\bf \method matches all specs}, while competitors
% miss one or more of the features.\label{tab:salesman}}
% \end{center}
% \end{table}

\section{Proposed Method}
\label{sec:meth}

\begin{figure}[ht]
    \begin{center}
        \begin{tabular}{c}
          \includegraphics[width=0.48\textwidth]{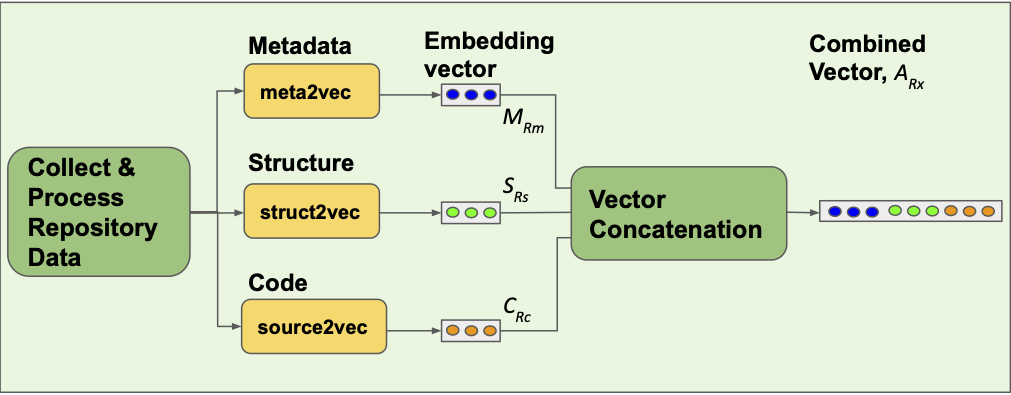} \\
        \end{tabular}
        \caption{Overview of the \method embedding: (a) we create an embedding representation for metadata, structure, and source-code, and (b) we combine them into an embedding that captures all three types of information. Each embedding hides significant subtleties and challenges.
          \label{fig:repo2vec} }
    \end{center}
\end{figure}

The main idea behind \method is to combine the  metadata, source code, and directory structure of a repository and provide an embedding representation for the whole repository. 
In fact, we create an embedding for each type of data,
which we refer to as:
 (i) meta2vec for metadata, (ii)  source2vec for the source code, and (iii) struct2vec for the directory structure.
 Our approach follows  these four steps.
 In the first three steps, we create an embedding vector for each of the three types of data, and in the fourth step, we combine these into a repository embedding.
 The \method pipeline is shown in Figure~\ref{fig:repo2vec}. We explain each step of our approach in more detail below.  

% \miii{This should be nearly a repetition of the novelty from the introduction, with maybe a bit more detail. However, using the same M dimensions should be a choice not a hardwired feature of the algorithm. Determining the right dimensions should be a paragraph somewhere in this section.}
 
 %\miii{Check if you need to use the remaining anywhere: (a) we create an M-dimensional meta vector, $M_{R_M}$, (b)  we create an M-dimensional structure vector, $SV_R$, (c)  we create an M-dimensional source code vector, $SV_R$, and (d) we combine these vectors into a single 3M-dimensional  vector, $RV$. }

\begin{figure}[t]
    \begin{center}
        \begin{tabular}{c}
          \includegraphics[width=0.47\textwidth]{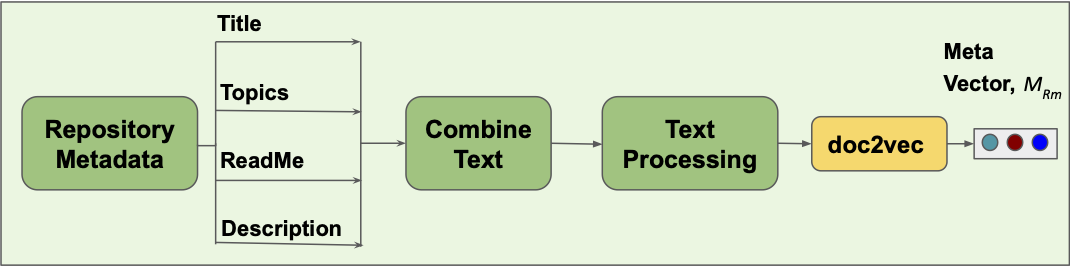} \\
        \end{tabular}
        \caption{The overview of the meta2vec embedding: (a) we collect the text from metadata fields, (b) we combine them into a single document, (c) we preprocess the text in the document, and (d) we map the document to a vector using an approach inspired by doc2vec.
          \label{fig:meta2vec} }
    \end{center}
\end{figure}

{\bf Step 1. Metadata embedding: meta2vec.}
We define meta2vec as mapping all the metadata 
%\miii{Use the same expressions consistently: learning, mapping, embedding, ... better use one}
in a repository to an $R_M$-dimensional embedding vector, $M_{R_M}$. In meta2vec, we follow three steps. First, we select the fields of metadata that we want to ``summarize" in embedding. Second, we preprocess the metadata text to remove noise. Finally, we adapt the doc2vec approach to compute the embedding vector. The overview of meta2vec is shown in Figure~\ref{fig:meta2vec}.
%\miii{Again: why not keep the exact dimension open: $R_x$ for each data type?}

{\bf a. Field selection: }
We consider all the fields of metadata that contain descriptive information regarding the content of a repository such as title, description, 
topics (or tags), and readme file. Recall that all this information is provided
by the author.
There are many ways to extract and combine textual information from each field.
Here, we opt to 
 treat  each metadata field as a paragraph and concatenate them to generate a document,
 which we process as below. 
 Note that we do not consider metrics that relates to the popularity of a repository, since our intuition and initial results suggest that it is less helpful in determining similarity.
 
% In the future, we intend to consider numeric fields such as stars, watches, and forks which capture the popularity of a repository but not the textual content.   

%However, these text are short and not well structured including special character, and other entity level noises.

{\bf b. Text preprocessing: }
Like any Natural Language Processing (NLP) method, we start with necessary preprocessing of the text to improve the effectiveness of our approach. 
%We need to have preprocessing step to eliminate the noise from the metadata mentioned above. 
As metadata in a repository text fields are often noisy, we follow the NLP best practices step which include removal of: (i) special characters e.g. `?', and `!', (ii) irrelevant words and numbers e.g. ``URL", ``Email", ``123", (ii) stopping words. 
%\miii{Rewrite more smoothly. It is very awkward now. I think we had a better phrasing in SFinder. Put emphasis on standard NLP practives, that's all you need to say, and start from that. Then you can just explain somethings. Note that \# could be useful as in C\#...}

{\bf c. Repository meta vector generation: }
We map 
%\miii{Yet another way to describe the mapping! Also is this vector continuous as opposed to non-continuous?} 
the metadata in a repository to an $R_M$-dimensional distributed vector, $M_{R_M}$ in this step. Following the basic principles of doc2vec~\cite{le2014distributed} approach, we adapt it to our needs and constraints here. Specifically, as metadata in a repository often consists of unstructured text and is small in size, we employ PV-DBOW, discussed in Section~\ref{sec:background}, because it performs better for small text dataset. 
%\miii{Maybe try to not make it sound that we just use it: We follow the basic principles of doc2vec, which we adopt for our needs and constraints here. Specifically, we....or Our approach is inspired by the principles of doc2vec.} 
%on the documents from the preprocessing step and compute the  metadata vector, $MV_R$ for each repository. As the repository metadata is often small in size, and unstructured, we use the PV-DBOW version of the doc2vec approach on the documents presented in Section~\ref{sec:background}. 
%\miii{Can you mention that BOW seemed to give comparable if not better results compared to the alternative? Extrapolating from SourceFinder without mentioning it specifically}
%We use the PV-DBOW instead of PV-DM to pick words from the document instead of surrounding words of a target word. 
%As the repository metadata often contains smaller text size and they are mostly unstructured, PV-DBOW performs better.
% Figure \ref{fig:tikz} and \ref{fig:tikz2} illustrates the intution
% \reminder{comment them out, if the tikz package is missing}

% \begin{figure}[htbp]
% \begin{centering}
%     % if tikz is missing, comment out all the lines below % 'tikz comment out'
%     \input{FIG-TIKZ/sample-fig-tikz.tex}
% \caption{Sample tikz figure \label{fig:tikz}}
% \end{centering}
% \end{figure}

% \begin{figure}[htbp]
% \begin{centering}
%     % if tikz is missing, comment out all the lines below % 'tikz comment out'
%     \input{FIG-TIKZ/sample-fig-tikz2.tex}
% \caption{Another Sample tikz figure \label{fig:tikz2}}
% \end{centering}
% \end{figure}

\begin{figure}[t]
    \begin{center}
        \begin{tabular}{c}
          \includegraphics[width=0.48\textwidth]{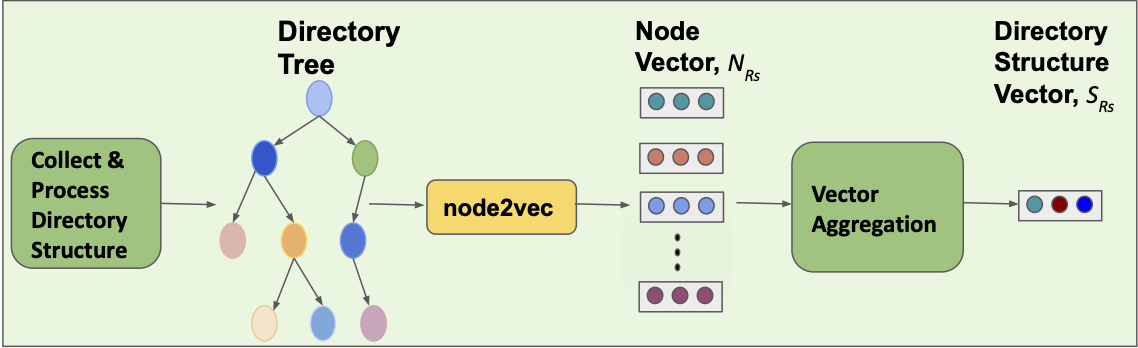} \\
        \end{tabular}
        \caption{The overview of our struct2vec embedding: (a) we extract directory tree structure of the repository, (b) we map each node into a vector following a node2vec approach, (c) we combine the node embedding to create the structure embedding for the repository.
           \label{fig:struct2vec} }
    \end{center}
\end{figure}

%\miii{This whole Step 2 needs rewriting to sound smooth: see what I did in the previous or my comments and repeat that here as well. First sentence in paragraph layes the overview and key idea. Then you explain steps in a smooth language.}
{\bf Step 2. Directory  structure embedding: struct2vec.}
We define struct2vec as mapping of repository directory structure to an $R_S$-dimensional embedding vector, $S_{R_S}$. We compute struct2vec following three steps. First, we represent the directory structure into a tree representation. Second, we generate node vectors employing node2vec. Third, we synthesize node vectors into a single structure vector. The overview of struct2vec is shown in Figure \ref{fig:struct2vec}.  
%utilize node2vec strategy in struct2vec, an embedding approach to represent a repository directory structure into an M-dimensional vector. In struct2vec, we follow the following three steps. First, we represent the directory structure into a tree representation. Second, we employ node2vec, presented in Section \ref{sec:background}, on this tree representation. This step computes an M-dimensional embedding vector for each node in the tree.  Finally, we aggregate these vectors column-wise into a single M-dimensional vector, $SV_R$ for a repository. The overview of our approach to compute struct2vec is shown in Figure \ref{fig:struct2vec}, and discussed below in details.

{\bf a. Directory tree representation: }
A software repository in \github consists of a standard directory structure with necessary data files and source code files. We consider the directory structure and transform it into Tree representation to enable node2vec on it. Note that, in order to nullify the effect of directory or file names in the mapping, the representation does not include directory or file names in the tree.

{\bf b. Node vector generation: }
We map all nodes in the tree into an $R_S$-dimensional node embedding vector, $N_{R_S}$, in this step. Following the properties of node2vec, first, we convert the trees into a set of paths using a biased random walk sampling strategy to include a diverse set of neighborhood nodes for a node. Then, we apply skip-gram models on these paths to get vectors for all nodes.   
%Second, the model computes embedding vector  for each node, $SV_N$ in the tree. These vectors are passed to next stage to perform vector aggregation to compute a single M-dimensional directory structure vector.

{\bf c. Repository directory structure vector generation: }
We compute repository directory structure embedding vector, $S_{R_S}$, by synthesizing the node vectors, $N_{R_S}$, in the tree. We follow column-wise aggregation method to synthesize these into a single vector.
%for all the nodes in the tree, we aggregate these vectors column-wise in a single vector, $SV_R$.
In order to do that, we employ six aggregation functions: mean, mode, max, min, sum, and standard deviation to compute a value for a column in the resultant vector.
%a single vector from a number of node vectors.  

\begin{figure*}[t]
    \begin{center}
        \begin{tabular}{c}
             \includegraphics[width=0.85\textwidth,height=6.97cm]{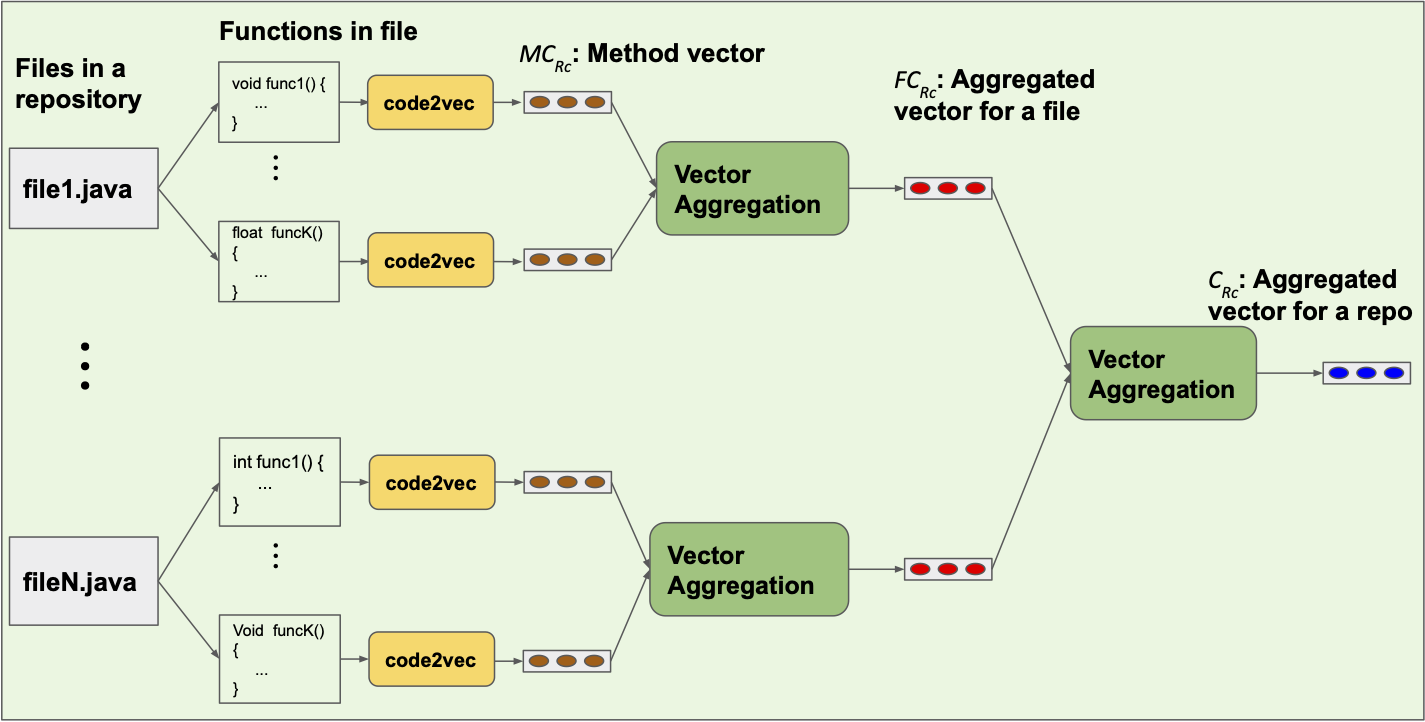} \\
        \end{tabular}
        \caption{The overview of our source2vec embedding: (a) we extract functions (methods) from each source file, (b) we embed each function, (c) we combine each function  embedding to create an embedding for each file, and (c) we aggregate each file embedding to create the source-code embedding for the repository.
           \label{fig:code2vec_pipeline} }
    \end{center}
\end{figure*}

{\bf Step 3. Source code embedding: source2vec.}
We define source2vec as an embedding approach to represent the source code in a repository to an $R_C$-dimensional embedding vector. In source2vec, we employ the Java method embedding techniques and a trained model with ~15.3M methods discussed in Section \ref{sec:background}. We follow three steps in source2vec. First,
we compute the $R_C$-dimensional method code vectors for each method in the source file available in a repository. Second, we aggregate these method vectors in a single $R_C$-dimensional file code vector. Finally, we compute the final $R_C$-dimensional repository code vector for all the source files by another level of vector aggregation.  The pipeline of our approach is shown in Figure \ref{fig:code2vec_pipeline} and discussed below in details.

% To create the embedding vector for source code present in a repository, we use the trained model with the java-large dataset with ~15.3M methods sourced from ~\cite{alon2019code2vec}. The model gives embedding vector only for a method. We employ two level of aggregations to get the repository embedding. First, we aggregate method vectors in a file to get the file vector. Later, we aggregate file vectors in a repository to a single repository vector using an aggregation function. 

{\bf a. Method code vector generation: }
A software repository may have multiple source code files and other files. First, each source file is decomposed into its methods. Next, methods are preprocessed into AST paths, and context vectors which are the input to the code2vec model. The model maps each method into an $R_C$-dimensional embedding code vector, $MC_{R_C}$. These method vectors are then passed to the next stage of pipeline to be aggregated into a single vector.

{\bf b. File code vector generation: }
After generating the method code vectors, $MC_{R_C}$, in a file, the task is now to aggregate them into an $R_C$-dimensional file code vector, $FC_{R_C}$. We follow a number of column-wise aggregation functions. The aggregation functions that we investigate are mean, mode, max, min, sum, and standard deviation. 
%Taking a column values for all method vectors, the function returns a value. 
Following the procedure, the pipeline creates a single file code vector, $FC_{R_C}$, and passes it to the next stage to create a single repository vector.

{\bf c. Repository code vector generation: }
At this stage of the pipeline, source2vec aggregates all the $R_C$-dimensional file code vectors, $FC_{R_C}$ for all source code files available in the repository to a single $R_C$-dimensional repository code vector, $C_{R_C}$. The pipeline follows same procedure like previous step, column-wise aggregation function to get the repository code vector.

% To create the embedding vector of repository directory, we use the trained model with the dataset, the methods sourced from ~\cite{grover2016node2vec}. The model analyzed the graph's structure and generate embedding vector for each node. Later we aggregate them into a single vector to get the directory embedding using a novel aggregation function.

% {\bf a. Repository Directory Vector Generation: }
% Repository directory’s structure varies from type to type. First, we generate tree structure of each repository. Next, get edgelist file for each tree, which contains node-edge pair information, and it will be used as the input of the node2vec model. The model infers a repository into a group of embedding directory vector for each node. Finally, we aggregate those group of vectors column-wise in a single vector, the generated vector will represent each repository. 

{\bf Step 4. \method: Repository embedding.}
% \begin{figure}
%     \begin{center}
%         \begin{tabular}{c}
%           \includegraphics[width=0.48\textwidth]{FIG/repo2vec.png} \\
%         \end{tabular}
%         \caption{Overview of the key functions of the Repo2Vec embedding: (a) we create an embedding representation for metadata, structure, and source-code, and (b) we aggregate them into an embedding that captures all three types of information. Each embedding hides significant subtleties and challenges.
%           \label{fig:repo2vec} }
%     \end{center}
% \end{figure}
We propose \method to present a \github repository in an embedding vectors using features from three types of information sources: metadata, source code, and project directory structure following the pipeline shown in Figure \ref{fig:repo2vec}. 
% First, we compute an M-dimensional vector, $MV_R$ from the metadata using meta2vec. 
% Second, we compute an M-dimensional vector, $SV_R$ from the repository structure using struct2vec. 
% Third, we compute an M-dimensional vector, $CV_R$ from the source code files using the source2vec mentioned earlier.
% \miii{This paragraph is a repetition here. Try to shrink. It sounds as if you are defining the method for the first time.}
In this step, we combine metadata vector $M_{R_M}$, directory structure vector $S_{R_S}$,  and source code vector $C_{R_C}$ into repository vector $A_{R_x}$. 

% \miii{Before the combination: {\bf Selecting the vector dimensions.}....Explain what and why. Discuss that we could have diff dimensions for each source, and we will consider it in the future... Provide arguments why having the same dimensions here kind of makes sense as it gives similar weight for each information type.}

Combining the vectors of each information type is a challenge as many methods exist following  two types of approaches: (a) merging the numerical values into a single vector, using the sum, average or median, etc, and  (b) concatenating  vectors to create a ``longer" vector. In both approaches one can consider weighting and normalizing to ensure ``fairness". 
Here, we opt to use the concatenation approach as follows:
%Finally, we combine these three vectors row-wise into a single 3M-dimensional vector, $RV$ by vector concatenation following the equation.  

\begin{equation}
A_{R_x} = w_M * M_{R_M} + w_S * S_{R_S} + w_C * C_{R_C}
\label{eq:weightedCombo}
\end{equation}

where $w_M, w_S$ and $w_C$ are the weights for the meta vector $M_{R_M}$, structure vector $S_{R_S}$, and source code vector $C_{R_C}$ respectively, and these weights are in the range of $[0,1]$. 

\section{Experiments and Evaluation}
\label{sec:exp}
We evaluate the effectiveness of \method using real data and answer two questions.

{\bf Q.1: What is the effect of each information type?}
We want to quantify the effect and contribution of the three information types in determining similarity.

{\bf Q.2: How does \method compare to prior art?}
We compare our method with
CrossSim~\cite{nguyen2020automated, nguyen2018crosssim},
which is arguably the state of the art
approach 
and was shown to outperform previous approaches~\cite{zhang2017detecting,mcmillan2012detecting,kawaguchi2006mudablue}.

% Here we conduct experiments to answer the following questions:

% \begin{compactenum}[{Q}1.]
% %\item  How effective is \method compared to the baseline approaches?
% \item How effective is \method compared to the CrossSim?

% \item  How does the metadata, source code and structure affect the performance?

% %\reminder{\item some other question, if any}
% % {\bf Accuracy}: How well can it reconstruct the actual values

% \end{compactenum}

\subsection{Experimental Setup}
We  present the datasets and our evaluation approach.
% the dataset, query set, ground truth, evaluation metrics, and similarity computation.

{\bf 1. Datasets.}
We consider two datasets in our evaluation: (a) 
a dataset of benign repositories, \dben, which
was used in prior work~\cite{nguyen2018crosssim, nguyen2020automated}, and (b) a dataset of malware repositories, \dmal, collected by a prior repository analysis study~\cite{rokon2020sourcefinder}. %\miii{We did not use a subset of SourceFinder? yes}. 
%  Using \github API, we crawl the metadata, source code, and the repository directory structures for all the repositories in datasets. 

%GUYS: we need a command for the dataset!

{\bf a. Benign repositories \dben:} This dataset consists of 580 Java repositories from \github  and was used in an earlier study introducing CrossSim~\cite{nguyen2018crosssim}. We select this dataset in order to make a fair and reproducible comparison with CrossSim. 
The dataset 
spans various software categories such as: PDF processors, JSON parsers, Object relational mapping projects, Spring MVC related tools, SPARQL and RDF, Selenium test, Elastic search, Spring MVC, Hadoop, and Music player.
% For the compatibility and reproducibility purposes, we collect the exact dataset used by CrossSim. 

% To compare our approach with the baseline study, CrossSim, we collect the same dataset with 580 java repositories from GitHub. The dataset covers repositories from various fields (e.g., PDF processors, JSON parsers, Object Relational Mapping projects, and Spring MVC related
% tools). Using GitHub API, we retrieve metadata (title, descriptions, topics, readme file), and  the source code files, and directory structure for all repositories. We call this dataset as {\em benign dataset} in the paper.
% \begin{table}[ht]
% \begin{center}
% \begin{tabular}
%   {|P{0.2\linewidth}|P{0.2\linewidth}||P{0.2\linewidth}|P{0.2\linewidth}|}
%   \hline
%   {\bf Dataset} & {\bf Size} & {\bf Query Set} & {\bf Size}  \\ \hline
%   D\_ben & 580 & Q\_ben & 50 \\ \hline
%   D\_mal & 433 & Q\_mal & 50  \\ 
%   \hline
% \end{tabular}
% \end{center}
% \caption{Dataset and Query Set. \miii{I don't think that query size is that critical to report. Why not report approximate number of distinct software families? In fact at the very least we could remove the Query set anyway, even if we don't add anything. it adds noise.}}
% \label{tab:dataset}
% \end{table}

{\bf b. Malware repositories \dmal:} This  dataset consists of 433 Java malware repositories. 
The dataset is provided by the SourceFinder project~\cite{rokon2020sourcefinder}, 
%\ofr{CITE-ourwebsite?**, will it give some kind of hint about the author? }
%\miii{Good point: I removed it, we can add the site in camera ready: \cite{sourcefinder}}
whose goal is to identify and provide  malware source code repositories.
Here, we choose only the Java language repositories, which are the focus of the CrossSim approach.
%The SourceFinder dataset includes 433 jave malware repositories, and we add 1657 repositories, by collecting all the forked repositories of the initial set of 433. The intention of this extension is twofold: (a) we increase the size of our dataset, and (b) it helps to generate ground truth for evaluation purposes, as we discuss below.
The repositories have a fairly wide coverage across malware families including: Botnets, Keyloggers, Viruses, Ransomware, DDoS, Spyware, Exploits, Spam, Malicious code injections, Backdoors, and Trojans.   

% We collect a malware repository dataset to asses our method on a real dataset. We collect a list of 433 java malware repositories from SourceFinder~\cite{rokon2020sourcefinder} which finds malware source code repositories in GitHub. To asses the variant of a malware repository, we intend to measure similarity among the forked repositories of the 433 original repositories. Thus, we crawl 2090 java malware repositories including 1657 forked repositories. We name this dataset as {\em malware dataset}.

%CHANGE D_malware everywhere! :-)

{\bf 2. Query-based evaluation.}
For consistency and fairness, we follow the evaluation methodology
and similarity metrics of prior work~\cite{nguyen2018crosssim}.
We conduct our evaluation by using similarity queries as follows:
a given repository, we want to identify its five most similar repositories. 
%\miii{The five most similar correct?}
% We choose two subset of repositories from \dmal and \dben respectively as our query sets. We then find the similar repositories of the query set repositories and evaluate the performance of \method in finding similarity repositories. 
% In more detail, we pick a subset of repositories from each dataset as a query set and call as {\bf Q\_benign} and {\bf Q\_malware}. 

{\em a. The query-set Q\_ben:}
For the sake of compatibility with CrossSim, {\em Q\_ben} consists of the same query set of 50 repositories as CrossSim. The query set spans various domains e.g. SPARQL and RDF, Selenium test, Elastic search, Spring MVC, Hadoop, and Music player. 
%As the dataset is not large, to include diverse domains, CrossSim did not choose the query set repositories randomly. 

{\em b. The query-set Q\_mal}: For the \dmal dataset,
we create a query-set by selecting 50 repositories uniformly at random. The query set includes various malware families such as Keylogger, Botnet, DDoS, Ransomware, Virus, Backdoor, Trojan, etc.
%\miii{Did we ensure that we get a representative from evere family or was it truly uniformly random selection?}
% To reduce the manual investigation, we include in {\em Q\_malware} only the original repositories which have more than five forks. 
%We consider the forked repositories as the ground truth to evaluate the similar repository finding capability.
%Note that, due to page constraint, we omit the query sets in the paper.

{\bf 3. Ground truth generation.}  We establish the groundtruth for each dataset by manual evaluation 
% To evaluate the similarity performance in \dmal, we establish a ground truth dataset in the absence of such. We leverage the forking property of \github to create the ground truth. Now, following the same evaluation strategy of CrossSim, we do not create any ground truth for \texttt{ D\_benign}. We incorporate manual investigation strategy to evaluate the similarity performance for \texttt{ D\_benign}.
%{\bf a. The ground truth for \dben.} We establish the quality of the similarity for benign repositories using manual evaluation. 
and follow the scoring framework, which was used in prior work~\cite{nguyen2018crosssim}.
Namely, we use four categories of scores to label the level of similarity: 
%which we denote as as {\bf Expert Similarity Score (ESS)}:
%\miii{We said this needs revisiting: is this she same levels as CrossSim used?} \ofr{Yes}
%\miii{The levels do not represent visually the similarity disassimilrity concept: ESS1 sounds are lower in similarity versus not similar. How about using something different? E.g. ESS4 becomes VSimilar, ESS3 is SimilarEnough, ES2 is neutral, ES1 is Disssimilar}according to the following definitions:

\begin{itemize}
\item Category 4: Strongly Similar ({\bf SS}) repositories. %with respect to the metadata, source code, and project directory structure. 

\item Category 3: Weakly Similar ({\bf WS}) repositories.
%i.e. in metadata and/or source code and/or project directory structure. 
\item Category 2:  Weakly Dissimilar ({\bf WD}) repositories. %\miii{neutral does not sound as dissimilar Why not call it as "mostly dissimilar" or something.}
%\item ESS 2: The repositories are not similar, but they are from the same family or domain. e.g. both repositories are botnets.
\item Category 1: Strongly Dissimilar ({\bf SD}) repositories.
%\item ESS 1: The repositories are not similar and with nothing in common.
\end{itemize}

For consistency, we follow the convention of the previous study~\cite{nguyen2018crosssim}: a repository in category 3 or 4 is considered (sufficiently) similar or a True Positive. Conversely, a repository in category of 1 or 2 is considered dissimilar or a False Positive.
%\miii{Needs revisiting also, }

For the evaluation, we opted to use experts, who are more reliable
compared to a Mechanical Turk platform for highly technical questions~\cite{gharibshah2020rest} .
%\miii{Joobin made claim like this in his PKDD paper I think.}
Specifically, we recruited three computer science researchers with at least 3 years of Java programming experience. 
%They were instructed to manually label the similarity score.  
The evaluators are given the target repository and the response of 5 repositories per query. 
Note that the five repositories in each response are in 
random order to avoid introducing biases. 
The evaluators assign a score among the four categories of scores to each repository in the responses.
%, which we refer to as Expert Similarity Score (ESS). 
The evaluators were provided with context and information in order to calibrate their criteria. 
The first and second evaluators independently assign a score to each repository in the response. 
Later, the third evaluator acts as judge by rechecking and finalizing their scores if their scores are not same for a query. 

{\bf 4. Evaluation metrics.}
For consistency, we adopt the metrics used in related works~\cite{nguyen2018crosssim}, which we describe below.
% These metrics are: (a) success rate, (b) precision, (c) confidence, and (d) ranking, which we explain below. 

{\bf a. Success rate:} We say an answer to a query
is successful, if one or more of the returned repositories is similar to the above definition of similarity.
The  success rate is the percentage of successful queries.
%\miii{FUTURE: This is something that in the future we can vary: success rate could be a function of the number of queries that need to be similar SR(k)!}

{\bf b. Precision:} Precision is the percentage of the returned repositories which are similar to their query repository. We compute the precision following the equation, 

\begin{equation}
precision = \frac{SS+WS}{SS+WS+WD+SD}    
\label{eq:precisionWeak}
\end{equation}

% Therefore, we calculate the proportion of number of manually labeled similar repositories (ESS 3 or 4) to total number of repositories (250 for  \dben returned by our evaluation method.

{\bf c. True and False Positives:} Following the standard definitions, True Positives for a query-set is the total number of similar repositories returned, while False Positives is the number of non-similar repositories in the answers.

% {\bf d. False Positives:} in We measure the number of True Positive (ESS 3 or 4) and False Positive (ESS 1 or 2) using the ESS for all the queries. 

{ \bf d. Ranking order correlation (ROC):} We quantify the quality of the ranked answer to the query using again a metric introduced in prior work.
The intuition is to "reward" an algorithm that returns highly similar repositories ranked higher.
% The correct answer should not only identify the right repositories but also rank them in the right order.
% ranking order correlation (ROC) between ordered ideal ranking and ordered ranking by manual labeling. The most similar repository identified by our approach should be labeled as top similar repository by manual evaluation as well. Similarly, the worst repository identified by tool should also be the worst by the human evaluation. 
To quantify this, we calculate the widely-used  Spearman's rank correlation coefficient $r$~\cite{spearman1961proof}, which is defined as: 
\begin{equation}
r = 1 - \frac{6\sum (d_i)^2}{n(n^2-1)}
\label{eq:ROC}
\end{equation}

where $r$ is the coefficient, $d_i$ is the difference between the two ranks of each repository, and $n$ is the number of ranked repositories. The coefficient is in the range of $[− 1,1]$, with 1 implying perfect agreement, and -1 disagreement between the two rankings. 
\miii{Therefore we need to explain: Answers are in order of most similar, and we want to assess the order as well.}

% We calculate the metric for all pairs of ranking given by the our method and manual evaluation. An ROC value of +1 means that they are highly positively correlated (ranking is good), and -1 means that they are highly negatively correlated (ranking is total opposite). 

Comment: Given the way we formulate the query,
the use of Recall is less relevant here:
we ask the algorithms to report only the top five most similar repositories.
Formulating a query we expect the methods to return 
all similar repositories is challenging for two reasons.
First, we would need an established ground-truth, since
manual validation would be labor-intensive.
Second, there is no absolute way to define what constitutes 
``sufficiently similar" repositories,
while relative similarity is easier to define.

% The deeper reason is that since the 
% identifying all similar repositories for a query repository in the absence of ground-truth is challenging.  We intend to investigate Recall in the future once we identify an appropriate benchmark dataset.\miii{Not sure where to best place this}

% We describe the above-mentioned metrics from the perspective of evaluating the performance for \dben. For evaluating the \dmal, we only use success rate and precision metrics. The similarity rankings are compared by taking the set intersection between similar repositories return by our evaluation method and the ground truth. More details about the metrics can be found in \cite{nguyen2018crosssim}.

\subsection{Deploying \method}

We implement our method, which we described in Section~\ref{sec:meth} 
 using Python3.6 packages: TensorFlow2.0.0, gensim PV-DBOW doc2vec.
We discuss some implementation details and parameter choices.
% to generate the final \method vector by concatenating the embedding vectors from three different information sources: metadata, structure, and source code in a repository. We describe some of the choices for generating the final \method vector below.

{\bf Selecting the embedding dimensions.} We select 128 as the embedding vector dimension for $R_M$, $R_S$, and $R_C$, since well-established embedding techniques~\cite{mikolov2013distributed, le2014distributed, alon2019code2vec, grover2016node2vec} recommend
this number for striking a balance  between computational cost and effectiveness. 
We use the same number of dimensions 
for the vector of each type of information for fairness. 
Concatenating these three vectors creates 
a single \method vector with $R_x$=384 dimensions. 
The above choices give good results as we will see later.
In the future, we will explore the effect of different vector dimensions. 

{\bf Exploring the solution space via weight selection.} 
The weights  in equation~\ref{eq:weightedCombo} give us the ability to control the "contribution" of each information type. 
Here, we focus on the following weight  combinations,
which give rise to three derivative algorithms:
 (a) {\bf  \meta} using only metadata with weights
 $w_M=1, w_S=0, w_C=0$;
  (b) {\bf \metanode} using metadata and structure with weights $w_M=1, w_S=1, w_C=0$;  and,
  (c) {\bf \metanodecode} using  all three types of information with weights $w_M=1, w_S=1, w_C=1$.

In other words, we explore the  effect  of  weights  but  in  a  coarse  way.  In  the future,  we  intend to  explore  non-integer weight  combinations.
Overall, our results suggest that equal weights seem to work quite well,
but a savvy user can customize them to achieve optimal performance
for niche problems. 

% We select 1.0 as the weight for concatenating meta2vec, struct2vec, and source2vec into Repo2Vec.
%We tried a combination of $w_M, w_S, w_C$ to assess the effect of the information types, and find that Repo2Vec\_All ($w_M=1, w_S=1, w_C=1$) outperforms Repo2Vec\_MS ($w_M=1, w_S=1, w_C=0$), and Repo2Vec\_M ($w_M=1, w_S=0, w_C=0$). As evaluation takes a huge manual evaluation, we opt out to check a range of weights value. In the future, we intend to investigate with varying weights for each type of information sources. 
%\miii{I CANOT BELIEVE THAT the exact value of w as long as it is the same has ANY effect on the results. Can you clarify and maybe even rephrase?}

{\bf Calculating the similarity.}
There are many different ways to calculate the
similarity in an embedding space as the inverse of their distance in that space.
Here, we use the widely used cosine similarity,
which is often recommended  for high dimensional spaces~\cite{sidorov2014soft}, and yields
great results here as well.

% It measures similarity calculating the cosine of the angle between two vectors, $\vec{a}$ and $ \vec{b}$ using their dot product: 
% %We calculate the similarity using the following equations. 

% %\[\vec{a}^{\,} . \vec{b}^{\,} = ||\vec{a}^{\,}|| \hspace{0.1cm} ||  \vec{b}^{\,}|| \hspace{0.1cm} cos\theta \]
% \[ cos\theta = \frac{\vec{a}^{\,} . \vec{b}^{\,}}{||\vec{a}^{\,}|| \hspace{0.1cm} ||  \vec{b}^{\,}||}    \]

% The cosine similarity  ranges from -1 to 1 with a value of  1 indicating highly similar repositories. 

% The cosine similarity between $\vec{a}^{\,}$ and  $\vec{b}^{\,}$ is shown in Figure \ref{fig:cosine_sim}. 

% \begin{figure}
%     \begin{center}
%         \begin{tabular}{c}
%           \includegraphics[width=0.45\textwidth]{FIG/vector_similarity.png} \\
%         \end{tabular}
%         \caption{cosine similarity of two vector $\vec{a}^{\,}$ and  $\vec{b}^{\,}$ (a) highly similar as the $\theta$ is close to 0, (b) neutral as the $\theta$ is close to $\pi/2$, and (c) highly dissimilar as $\theta$ is close to $\pi$.
%           \label{fig:cosine_sim} }
%     \end{center}
% \end{figure}

 %We select 128 as the embedding dimension as well established embedding techniques~\cite{mikolov2013distributed, le2014distributed, alon2019code2vec, grover2016node2vec} suggest 128 is the optimum number of dimension with respect to computation cost and effectiveness. 
%Second, we  find that 128 is lowest dimension that empirically exhibits better performance.  
%We argue that setting 128 for each type of information give the equal representation for fairness.

{\bf Selecting the right aggregation function to aggregate multiple vectors into a single vector.}
As we see in Section \ref{sec:meth}, we introduce six column-wise aggregation functions to aggregate vectors into a single vector. We find that mean aggregation function performs better than others.
%which finds the mean value for a set of values of a column of all given vectors and return it as aggregated value for a specific column. 
In more detail, we evaluate the performance of all aggregation functions: average, max, min, mode, sum, and standard deviation. We find that embedding with mean aggregation shows highest 93\% precision for \dben dataset and 95\% precision for \dmal dataset. Max aggregation function shows the second best result 88\% and 91\% precision for benign and malware dataset respectively. Other aggregation functions show relatively lower precision for both dataset. In the remaining of the work, we use the mean aggregation function.

\subsection{Evaluation}
We evaluate \method in two ways. First, we assess the effect of each type of information
%, meta2vec, struct2vec, and source2vec
on the performance. Second, we compare our method against CrossSim~\cite{nguyen2018crosssim}, which is the state of the art approach.

%{ \bf \em Which configuration works best in computing similarity between repositories?}
{\bf a. The effect of the information types:}
We evaluate the effect of information types by comparing the
performance of our three variants: \meta, \metanode, and \metanodecode, which
we defined earlier.
We report the result in Table~\ref{tab:comp_model_config} for our three \method variations and both datasets.
This evaluation leads to two main observations:

\begin{table}[t]
\begin{center}
%\footnotesize
\begin{tabular}
%{|p{0.10\linewidth}|r|r|r|r|r|r|r|}
  {|P{0.2\linewidth}|P{0.14\linewidth}|P{0.14\linewidth}|P{0.14\linewidth}|P{0.14\linewidth}|}
  \hline 
   \cellcolor{blue!30} & \multicolumn{2}{|c|}{\cellcolor{orange!50!}\textbf{\dben Dataset}} & \multicolumn{2}{|c|}{\cellcolor{orange!50!}\textbf{\dmal Dataset}} \\
    \cellcolor{blue!30} \textbf{Method} & \cellcolor{green!20} Success Rate & \cellcolor{blue!30!green!30!} Precision & \cellcolor{green!20} Success Rate & \cellcolor{blue!30!green!30!} Precision \\ \hline
  Repo2Vec\_M & 100\% & 62\%	& 100\%	 & 67\% \\ \hline
  Repo2Vec\_MS & 100\% & 76\%	 & 100\% & 82\% \\ \hline
  Repo2Vec\_All & {\bf 100\%} & {\bf 93\%}	 & {\bf 100\%} & {\bf 95\%}\\
  \hline
\end{tabular}
\end{center}
\caption{ Performance comparison of our three variants of \method. Using all three information types (metadata, structure, and source code) provides significantly better results.}
\label{tab:comp_model_config}
\end{table}

    {\bf  Observation 1: Using all three data types provides significantly better performance.} 
    In the table, we see that \metanodecode achieves 93\% and 95\% precision compared to 76\% and 82\% when only metada and structure information are used.
    
     {\bf Observation 2: Metadata and structure provide fairly good results.} Although \metanodecode performs best, \metanode performs quite well especially if we compare it with CrossSim on the same benign dataset and query-set shown in Table~\ref{tab:model_comp}. Note that the computational effort for using metadata and structure is significantly less compared to analyzing the code.
     \miii{Optional discussion: From a practical point of view, the result suggests that the metadata and structure together provide a reasonable amount of information. Thus, for example, 
       if we were doing manual analysis examining the metadata
       and the structure, could provide a good basis for a classification.
       NOT SURE as this can open up discussions...
     }
     
    % Structural information complements metadata, and increase the precision from 62\% to 76\%, and 67\% to 82\%. Later, adding source code information improves the precision from 76\% to 93\%, and 82\% to 95\% since source2vec captures semantic meaning of program.

\miii{FUTURE: We could add Discussion. on the balance between computation effort and accuracy. }

% Our evaluation shows that \metanodecode exhibits the best performance in finding similar repositories but only \metanode mis enough to produce a competitive result.

% In details, we consider two query sets, {\em Q\_benign}  with 50 benign repositories, and {\em Q\_malware} with 47 malware repositories. First, our approach computes the top 5 similar repositories for each query applying cosine similarity on a pair of repository embedding vectors. Second, the results for benign query set is manually investigated and labeled with ESS. However, the results for malware query set is tested against the ground truth. Third, we compute the success rate and precision to evaluate \method. 
% The results are presented in Table \ref{tab:comp_model_config}. Here, we see that the \meta created from only metadata achieves  100\% success rate and 62\% precision for the \dben dataset, and 100\% success rate, and 80\% precision for the \dmal dataset.  However, the precision is relatively lower than other two approaches. In terms of precision, \metanode embedding with the metadata and structure data, performs better than \meta, and \metanodecode embedding with the metadata, structure, and source code shows the best precision result for both dataset. Analyzing the result, we find that embedding with only metadata and structure is enough to find similar repositories, and competitive to the state-of-the-art techniques. However, \metanodecode includes information from three types of information sources of a repository which helps finding similar repositories precisely.   

\begin{figure}[t]
    \begin{center}
        \begin{tabular}{c}
           \includegraphics[width=0.45\textwidth,height=5.5cm]{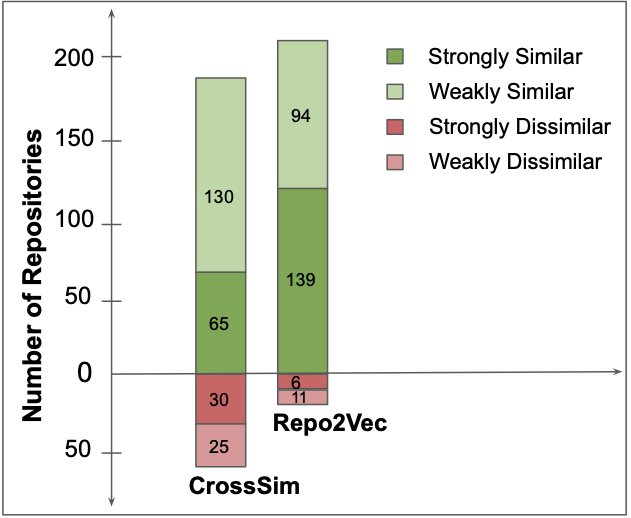} \\
        \end{tabular}
        \caption{\method outperforms CrossSim significantly: it finds 
         nearly twice as many Strongly Similar repositories and 30\% fewer False Positives.
           \label{fig:confidence_compare} }
    \end{center}
\end{figure}

{\bf b. Comparing \method  to the state of the art.}
We compare the best configuration, \metanodecode, with  CrossSim with respect to success rate, precision, confidence, and ROC for the benign dataset \dben. We find that \method outperforms CrossSim in terms of precision and ROC and has the same  success rate as CrossSim.

\begin{table}[t]
\begin{center}
\begin{tabular}{|P{0.2\linewidth}|P{0.2\linewidth}|P{0.2\linewidth}|P{0.2\linewidth}|}
  \hline
  \cellcolor{blue!30} {\bf Method} & \cellcolor{green!20} {\bf Success Rate} & \cellcolor{blue!30!green!30!} {\bf Precision} & \cellcolor{blue!50!green!50!} {\bf Spearman's Coefficient (r)}  \\ \hline
  CrossSim & 100\% & 78\% & 0.23 \\ \hline
  Repo2Vec\_All & 100\% & 93\% & 0.59 \\ 
  \hline
\end{tabular}
\end{center}
\caption{\method performs better in comparison of similarity approaches between Repo2Vec and CrossSim for the \dben dataset.}
\label{tab:model_comp}
\end{table}

{\bf Observation 3: \method: higher precision and better ranking.}
The results are presented in 
Table~\ref{tab:model_comp}.
\miii{We need to define/explain how is precision calculated: using the SS and WS as True Positives, and using SD and WD as ne definition of similarity. I don't think we need to define strict similarity, we can call it a deeper dive into quality! }
Although {\bf CrossSim} does well in terms of success rate, its precision of 78\% is significantly lower compared to the precision of 93\% of \metanodecode. Also, the ranking of similar repositories identified by \metanodecode is better than {\bf CrossSim}. 
%We compute the ranking correlations of similarity score given by \metanodecode and manual evaluation using Spearman's rank correlation coefficient. 
%As the ranks identified by \metanodecode tool is in descending order, the lower coefficient means better correlated. 
We find that ROC = 0.59 for \metanodecode, and ROC = 0.23 for {\bf CrossSim}, which further suggests that \metanodecode is better at computing similarity among repositories.

% ARE YOU HERE? Now I am here - my daughter delays so I ahve some more time! Great for me
% you will add the fugyre rfom skype right?Yes

%{\bf \method: higher evaluator confidence.}
{\bf Observation 4: \method provides better quality results.}
%How confident are the evaluators in their labelling?
Given that we have four categories of similarity, we assess the quality of the results as follows.
We plot the returned repositories from each method per
category in Figure~\ref{fig:confidence_compare}. 
Considering category 4 (strong similarity) only, \metanodecode identifies nearly 100\% more such repositories!
Similarly,  CrossSim reports 5 times more repositories 
in the strong dissimilarity category.

%Considering categories 4 and 3, \metanodecode identifies

% --- old delete or maybe repuprose some text if needed ---
% The plot demonstrates the comparison of confidence scores: (a) repositories in levels 1 and 2 are False Positives, (b) repositories in levels 3 and 4 are True Positives.
% (a) True Positives: total \# repositories with ESS value 3 and 4, and (b) False Positive: total \# repositories with  ESS value 1 and 2. 
% %, \metanodecode shows better confidence than {\bf CrossSim} in identifying similar repositories. 
% We find that \metanodecode identifies only {\em 17 False Positive} and {\em 233 True Positive} where CrossSim identifies 55 False Positive and 195 True Positive. 
% Furthermore, we see that \metanodecode has a significantly higher number of repositories with an ESS score of 4.
% ---- end old text ---

In conclusion, 
our comparison
suggests that \method outperforms  CrossSim.
The evaluation is summarized in Table \ref{tab:model_comp} and Figure \ref{fig:confidence_compare}.
In addition, CrossSim was shown to perform better than other related works  RepoPal,  CLAN, and  MUDABLUE~\cite{nguyen2018crosssim}. 

\section{Case Studies}
\label{sec:application}
In this section, we want to showcase how 
%Can \method extract useful information from a real dataset?
\method can facilitate repository mining studies
for specific applications considering both unsupervised and supervised techniques. %useful information from a real dataset? 
We consider two likely case studies: a) classifying repositories as benign or malicious, and b) clustering a set of repositories.

\subsection{Identifying malware repositories}
We showcase the usefulness of our \method in a supervised classification problem, which is of interest to practitioners~\cite{rokon2020sourcefinder, soll2017classifyhub, zhang2019higitclass}. 
%\miii{Maybe add some more citations?}. 
The question is to identify whether a repository contains malware or benign code. 
We assess the effectiveness of our approach and we also compare it with the state of the art method~\cite{rokon2020sourcefinder}.
% We apply \method to see effectiveness of the embedding in malware classification.
% In order to do that,

We create a dataset of 580 benign repositories
from \dben and 433  malware repositories from \dmal collected and discussed in Section \ref{sec:exp}.

\begin{table}[h]
\begin{center}
\begin{tabular}{|P{0.15\linewidth}|P{0.15\linewidth}|P{0.15\linewidth}|P{0.15\linewidth}|P{0.15\linewidth}|}
  \hline
  \cellcolor{blue!30} \textbf{Method} & \cellcolor{green!20} \textbf{Accuracy} & \cellcolor{blue!30!green!30!} \textbf{Precision} & \cellcolor{blue!30!green!30!yellow!30!} \textbf{Recall} & \cellcolor{blue!50!green!50!} \textbf{F1 Score} \\ \hline
  SourceFinder & 90\% & 89\% & 99\% & 94\% \\ \hline
  Repo2Vec & 97\% & 98\% & 96\% & 97\% \\ \hline
\end{tabular}
\end{center}
\caption{\method outperforms SourceFinder in malware repository classification}
\label{tab:mal_classificaiton}
\end{table}

%\miii{Have we given it a name or we just refer to the general datasets? Maybe say without considering any of the forked repositories? Or do we random sampling?}

Using our \method, we determine the embedding vector for each repository.
For the classification, one can use a plethora of ML approaches. Here,
we use the Naive Bayes, which is widely used for NLP classification problems~\cite{xu2018bayesian},
and, more importantly, it is also used by the most recent
SourceFinder study~\cite{rokon2020sourcefinder}. 
With this selection, we want to focus more on the effect of the features
when comparing to the SourceFinder classification. We implement the 
SourceFinder classifier, and apply it on our dataset.

We assess the classification performance using 10-fold cross validation. The results are shown in Table \ref{tab:mal_classificaiton}. Our model  classifies the malware and benign repositories with 98\% precision and 96\% recall which clearly outperforms the previous malware repository classification study by SourceFinder~\cite{rokon2020sourcefinder}.  

%\miii{The reviewers will say: could you apply SourceFinder to the same dataset! I think we can and then we don't ahve to "compare results from different datasets", which is usually not easy to defend!}

\subsection{Hierarchical clustering}

\begin{figure}[t]
    \begin{center}
        \begin{tabular}{c}
           \includegraphics[width=0.45\textwidth]{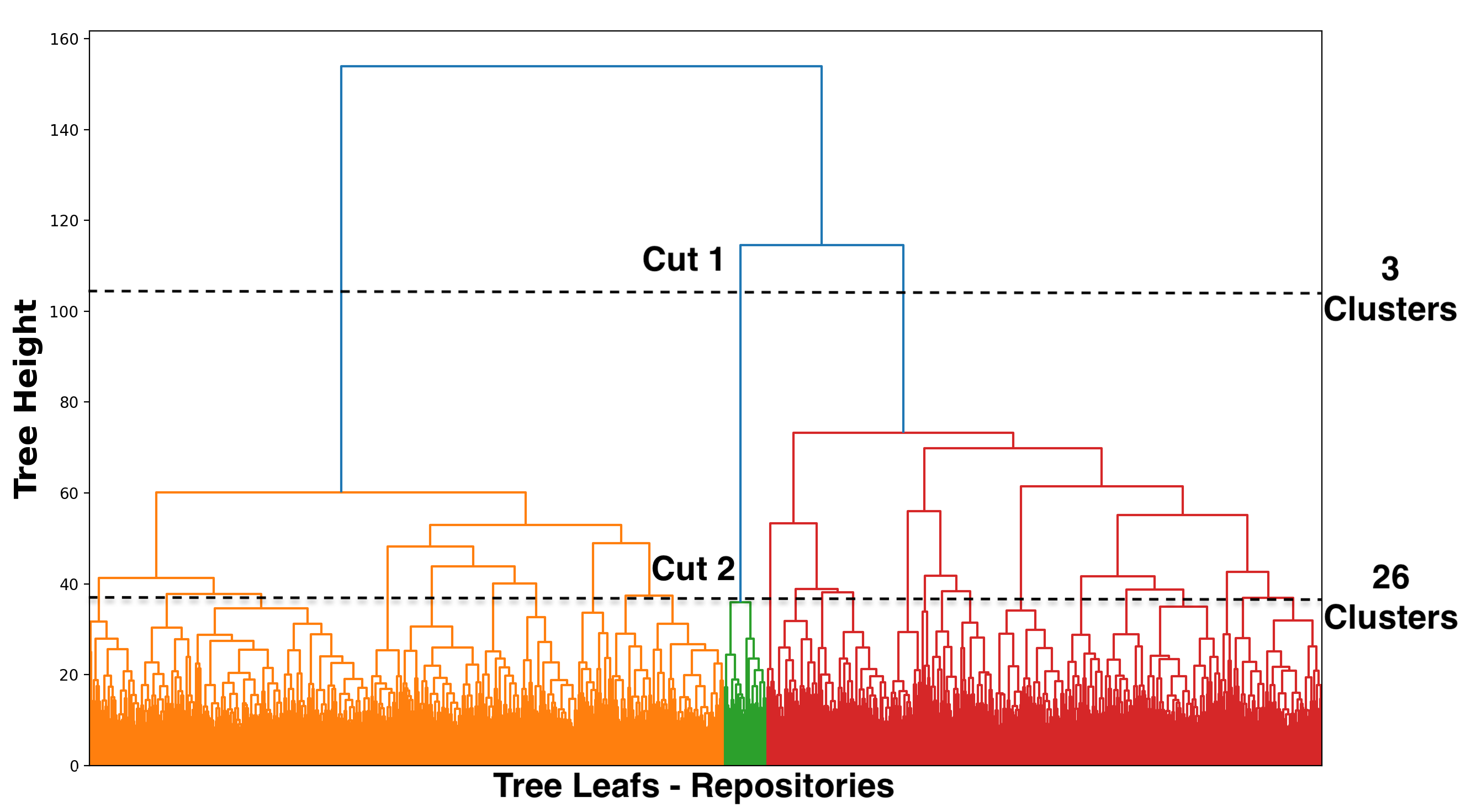} \\
        \end{tabular}
        \caption{Hierarchical clustering of malware repositories. Horizontal line 1 cuts into 3 distinct cluster of repositories and line 2 cuts into 26 distinct cluster of repositories
           \label{fig:dendogram} }
    \end{center}
\end{figure}

\begin{table}[t]
\begin{center}
\begin{tabular}{|p{0.08\linewidth}|p{0.1\linewidth}|p{0.16\linewidth}|||p{0.08\linewidth}|p{0.1\linewidth}|p{0.16\linewidth}|}
    \hline
    \cellcolor{green!20!yellow!60!blue!30} \textbf{Cluster No.} & \cellcolor{green!70!yellow!20!} \textbf{Number of Repos} & \cellcolor{blue!20!green!20!} \textbf{Dominant Repo Family} & \cellcolor{green!20!yellow!60!blue!30} \textbf{Cluster No.} & \cellcolor{green!70!yellow!20!} \textbf{Number of Repos} & \cellcolor{blue!20!green!20!} \textbf{Dominant Repo Family} \\ \hline
    \cellcolor{red!40}1 & 25 & DDoS & \cellcolor{red!40} 14 & 10 &   Virus \\ \hline
    \cellcolor{red!40}2 & 27 &   Android Keylogger & \cellcolor{red!40} 15 & 58 &   Trojan and Spyware \\ \hline
    \cellcolor{red!40}3 & 42 &   Backdoor & \cellcolor{green!40} 16 & 33 &  REST API \\ \hline
    \cellcolor{red!40}4 & 32 &   Worms & \cellcolor{orange!40} 17 & 48 & Hadoop \\ \hline
    \cellcolor{red!40}5 & 44 &   Android Botnet & \cellcolor{orange!40} 18 & 36 & JSON Parser \\ \hline
    \cellcolor{red!40}6 & 55 &   Android Malware & \cellcolor{orange!40} 19 & 45 & Music Player \\ \hline
    \cellcolor{red!40}7 & 31 &   Rootkit & \cellcolor{orange!40} 20 & 71 & SPARQL \\ \hline
    \cellcolor{red!40}8 & 24 &   Java Keylogger & \cellcolor{orange!40} 21 & 146 & Elastic Search \\ \hline
    \cellcolor{red!40}9 & 32 &   Ransomware & \cellcolor{orange!40} 22 & 54 & Object Relational Mapping \\ \hline
    \cellcolor{red!40}10 & 24 &  Whitehat Hacking & \cellcolor{orange!40} 23 & 27 & PDF Processor \\ \hline
    \cellcolor{red!40}11 & 15 &  Malicious Code Injection & \cellcolor{orange!40} 24 & 25 & Graph-Aided Searc \\ \hline
    \cellcolor{red!40}12 & 8 &  Android Trojan & \cellcolor{orange!40} 25 & 31 & Selenium   \\ \hline
    \cellcolor{red!40}13 & 6 &   Android Backdoor & \cellcolor{orange!40} 26 & 56 & Spring MVC \\ \hline
\end{tabular}
\end{center}
\caption{Fine-level clustering: the profile of the 26 repository clusters using a topic extraction method. The color of the cluster is similar to that of Figure~\ref{fig:dendogram}.}
\label{tab:20_cluster}
\end{table}

\begin{table}[t]
\begin{center}
\begin{tabular}{|p{0.1\linewidth}|p{0.1\linewidth}|p{0.13\linewidth}|p{0.47\linewidth}|}
    \hline
    \cellcolor{green!20!yellow!60!blue!30} \textbf{Cluster No.} & \cellcolor{green!70!yellow!20!} \textbf{Number of Repos} & \cellcolor{blue!20!green!20!} \textbf{Cluster Type} & \cellcolor{blue!20!green!20!yellow!20!} \textbf{Cluster Description} \\ \hline
    \cellcolor{red!40}1 & 433 & Malware & The \dmal  malware repositories \\ \hline
    \cellcolor{green!40}2 & 33 & Benign & Cluster 16 from the fine granularity with REST API repositories \\ \hline
    \cellcolor{orange!40}3 & 547 & Benign & The \dben repositories. \\ \hline
\end{tabular}
\end{center}
\caption{Coarse-level clustering: the profile of the three clusters.
The color of the cluster is similar to that of Figure~\ref{fig:dendogram}}.
\label{tab:4_cluster}
\end{table}

Here we showcase whether our approach can lead to a meaningful clustering of  repositories creating the basis for an unsupervised solution. We consider the union of our two datasets, \dmal and \dben dataset with a total of 1013 repositories.
%Without the use of an embedding, one would have to calculate all pair wise similarity scores with $O(N^2)$ cost in the number $N$ of repositories, which for large $N$ it becomes computationally expensive.
%\miii{Rokon: does the above make sense or is it overselling?}
%\miii{DMal repository and consider XXXX (all or some, if some which ones) repositories}.

First, we apply \method on all the repositories and get the embedding vectors. Second,
we apply the widely-used agglomerative hierarchical clustering (AGNES)~\cite{murtagh2014ward} on the  vectors of the repositories. 
Clearly, there are many different clustering techniques,
but note that our goal is to showcase the capability and
not to propose a clustering method.
%We use the euclidean distance as a measure of similarity between two vectors and we obtain a hierarchical clustering. 
We show the resulting hierarchical clustering in Figure~\ref{fig:dendogram}.

{\bf How meaningful is this clustering?}
Assessing the effectiveness of a hierarchical clustering is challenging
and it can depend on specific focus of a study.
A related question is at what levels of granularity  we should focus.
We provide indirect proof that our clustering provides meaningful results.

{\bf Considering two levels of granularity.}
We  analyze our hierarchical clustering at two different levels of granularity,
which are represented by two horizontal lines in Figure~\ref{fig:dendogram}. The
first line (Cut 1) corresponds to a {\bf coarse level} of granularity and yields
three large clusters. The
 second line (Cut 2) corresponds to  {\bf fine level} of granularity
 and yields  26 smaller clusters. 
% % \miii{We may want to think of different names, coarse vs fine level of granularity. Also it would be good to explain how we picked these levels.}
%  These two levels of granularity provide a view into levels of granularity in a way that it is still tractable for analysis.

\begin{figure}[t]
    \begin{center}
        \begin{tabular}{c}
           \includegraphics[width=0.45\textwidth, height=4.5cm]{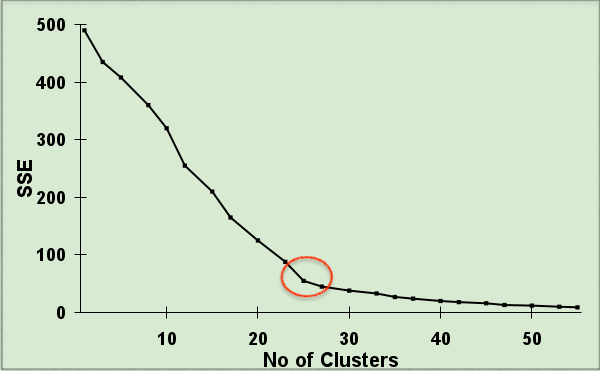} \\
        \end{tabular}
        \caption{ Determining optimal number of clusters. Diminishing returns of sum of squared error (SSE) is shown at red circle. 
           \label{fig:elbow} }
    \end{center}
\end{figure}

 We elaborate on how we select the two cuts in the dendogram in Figure~\ref{fig:dendogram}. 
  First, we select Cut 1 to see if the clustering distinguishes the malware from the benign repositories.  Second, we select a Cut 2 in a way that optimizes the number of clusters. A commonly-used approach is  the elbow method~\cite{kodinariya2013review}. The elbow or knee of a curve is a cutoff point in the number of clusters versus sum of squared error (SSE) graph, where increasing the number of cluster shows diminishing returns. Figure~\ref{fig:elbow} shows that the elbow lies at around K=26 clusters,
  which is how we select Cut 2.

Our goal is to profile the identified clusters at both levels of granularity.
%We select randomly 50\% of all repositories in each cluster  and profile them using manual inspection.
%We randomly select 50\% of all repositories in each cluster and profile them. In detail, first, we identify five most important keywords in metadata using LDA topic modeling~\cite{blei2003latent}. Second, we manually investigate the repositories to verify the accuracy of the extracted topics.
%\miii{Is it accurate to say a combination of automated tools (word count, and manual inspection? -- we don't want to lie.}
The results are shown in Table~\ref{tab:20_cluster} 
and~\ref{tab:4_cluster}. 

{\bf 1. Fine level cluster profiling:}
% In order to get details of larger cluster, we draw the second horizontal line in Figure \ref{fig:dendogram}. It cuts the tree into 20 smaller branches. As we can see from the figure, the previously mentioned largest cluster by line 1 is now divided into 10 smaller size sub clusters.
We want to evaluate the nature and the cohesiveness of the 26 clusters
at this level. 
We extract the profile of each cluster in terms of its focus
and we present the results in Table~\ref{tab:20_cluster}. 
Our profiling consists of two steps: (a) we identify the dominant keywords of the cluster and (b) we assess how aligned its repository is to the profile cluster.
In more detail, 
we identify the cluster topics using Latent Dirichlet Analysis (LDA) topic modeling~\cite{blei2003latent} on
the metadata of each repository. 
Note that we use a randomly selected subset of half of the repositories in the cluster.
Second, we want to identify the most dominant topic among all the candidate topics.
The most dominant topic is the one that appears in the most repositories of the cluster.
We report that topic in table~\ref{tab:20_cluster}.
The cohesiveness of the cluster is substantial:
at least 80\% of the inspected repositories are clearly members of the family of the cluster.
Finally, as an extra optional step, 
we manually investigate the repositories to verify the accuracy of the profile.
\miii{I don't think we need this last statement: presumably when we check dominance, we have already found this 80\%}

%As expected, each cluster is more ``specific". 
This process gives us both cohesive and "focused" clusters.
Most of the clusters contain repositories from  narrowly-defined malware or benign software families, such as Android Botnet, Keyloggers, Trojan, DDoS, Backdoor, Hadoop, Json parser, Elastic Search, and Spring MVC. 

We provide an indication of an insight that can be extracted here.
Interestingly, the largest malware cluster (cluster 15) with 58 repositories contains repositories from Trojan and spyware malware families. A Trojan malware program is similar to spyware except that it is packaged as another program.
This observation can give rise to the following hypothesis: could Trojan and Spyware have more in common than we thought?
\miii{This is an interesting direction, but as is, it is a bit weak. Maybe phrased as: This kind of unsupervised clustering can potentially provide interesting insights...
However unless you can close the loop and say that Trojan and Spyware are closer than Workds and backdoor...}

{\bf 2. Coarse level cluster profiling:}
% As we can see from the tree hierarchy Figure \ref{fig:dendogram}, the line 1 divides the tree into 4 large group of clusters  where similar repositories are grouped together. 
The overarching observation is that the three clusters of this level
correspond correctly to different
software domains as shown in Table~\ref{tab:4_cluster}. We find that following clusters: 
(a) the  \dmal, malware repositories, (b) the \dben, benign repositories,
and (c) REST API related benign repositories, which correspond to cluster 16
%\miii{define command: \rest}  
in the fine granularity clustering.
% First 3 clusters are fairly ``well defined" with around 97\% accurate decomposition where only 33 benign repositories are misplaced. 
% \miii{Not correct: our goal was not to detect malware but to cluster similar! Lets remove it}
The fact that the unsupervised clustering separated malware and benign repositories
suggests that malware and benign software are different.
%This separation comes as a bonus here, as it is not
The only exception seems to be
the REST API cluster 16\miii{Command for the 16 number!},
which would have been bundled with the malware repositories if we have created a two cluster decomposition.
We argue that the REST API repositories
 seem to resemble ddos and botnet malware (opening and listening to ports etc).
\miii{Can you provide an example that will viscerally explain: tool X, does XZY -keywords in meta, structure or functions,
which is reminiscent of XYZ in DDoS/Trojan/botnet malware}
%at least 80\% of the inspected repositories are clearly members of the malware of the family of the cluster.
%\miii{Can we argue that the first 3 clusters are fairly "well defined", ie at least 80xxxx\% of the inspected repositories were clearly members of the malware of the family of the cluster?   Right now, there is no such comment....}
% Note that the emergence of the larger and more diverse cluster is simply an indication that this cluster needs to be decomposed into more sub-clusters, which is what we do with our fine level clustering below.

\section{Discussion}
\label{sec:discuss}
In this section, we discuss the scope, extensions, and limitation of our study. 

{\bf a. What are the limitations of \method? } As \method is a comprehensive approach with data from three different sources, it performs even if every data source is not present. However, we believe unstructured software repositories with evasive metadata and obfuscated source code might fool Repo2Vec. In this case, previous works might perform better as these mostly depend on the graph connection of repositories.

{\bf b. Will our approach generalize to other programming language repositories?} 
Our approach is generalizable and extendable for all programming languages, though accuracy levels may vary. First,  
code2vec~\cite{alon2019code2vec}  can be extended to other programming languages, and the researchers seem to have plans to expand to other languages.
%\miii{THe below I don't understand... If code2vec works with other languages already or they plan to extend it, we can just refer it as such: 
Second, 
two information types, metadata, and structure, are fairly programming-language independent. 
Furthermore, from Table \ref{tab:comp_model_config}, we can see that even using
only these two information types, we can achieve reasonably good performance.
%}
%To generate source2vec, we use code2vec model~\cite{alon2019code2vec} which is available for java programming language source code. However, it is extendable for other programming language source code with a large set of dataset. For the lack of a large dataset, we limit our source2vec to work with the source code written in Java programming language.   

{\bf c. What will happen if the quality of metadata is low or misleading?} 
If metadata becomes unreliable, we could decrease its weight in our algorithm.
% Providing metadata is optional, and there is no check on the consistency or quality of the data. 
At the same time, 
we find that developers have an inherent motivation to provide quality metadata.
First, these repositories are part of the developers professional persona, and part of one's professional portfolio or resume. 
Second, these repositories are public, therefore there is an intention to make them
both easy to find and easy to use. The bragging rights of having a popular repository
is a strong motivation to provide informative metadata.
%\miii{If we can add statistics that: The below statistics provide indirect evidence of the above.   a) few empty metadata, b) mention the importance/perks of being a famous repository owner: e.g. Github user XXXX claims in her website that they own 10 repositories with more than XXX forks .}
Hence, the number of these type of repositories tend to be very low. We only have 1 in 580 (0.17\%) repositories in \dben, and 3 in 433 (0.69\%) repositories in \dmal with an empty metadata. 
Also, as Repo2Vec is a comprehensive approach with data from three information sources, even if metadata is unavailable, it will perform sufficiently.

%\ofr{1 in 580, 138 in 2090 (6\%)}
% we believe our approach will work fairly even if the quality of metadata is low for two reasons. First, our approach works by combining information from three different types of data sources. It will still be handy with other two embeddings. 

{\bf d. Why is \github search not sufficient to identify similar repositories?}
\github only allows the retrieval of repositories based on the keywords.
Though very useful, \github's query capability is not answering the problem that we address here.
%The service is not a 
% GitHub  offers  an  option  to  search  repositories, issues,  and  user  name  related  to  a  query. 
% However, it is not sufficient finding relevant repositories for three reasons.
First, it does not support query by example: "find the most similar repositories to this repository". 
Second, it does not provide the ability to measure similarity between a pair of repositories or  rank a group of repositories based on similarity to a given repository.
% Third, \github  cannot limit its search to an arbitrary subset of repositories of interest. 
%Fourth, 
Third, the service does not seem to use source-code which as we saw, provides significant improvement.  
%----------------
%\miii{SIGNIFICANT rewrite is needed: a) if GitHub has a function, we should compare if possible or claim that we cannot "focus" it on a given dataset, as we don't know how and Github looks for. b) it does not use source code (which we showed here that it provides significant improvement), c) it does not provide the ability to measure similarity between say a pair or get a ranking among a group of repositories. Finally,  can we argue that there are limitations on what and how we can queary? we CANNOT SAY that metadata is inaccurate! We just argued above it is likely to be accurate!}
% GitHub  offers  an  option  to  search  repositories, issues,  and  user  name  related  to  a  query. It  matches text present in the repositories and retrieves the relevant repositories. As providing text in a repository is optional for developer, it is not ideal to find similar repositories. The  text  might  also  be  irrelevant  to  project  as  GitHub does not impose any restrictions. Moreover, it does not care  the  actual  contents  of  the  repositories  like  source
% files and project structure. 

{\bf e. Are our datasets representative?} This is the typical hard question to answer for any measurement study.  We attempt to answer the question by making two statements. First, we evaluate our approach with the same dataset of 580 repositories (\dben) used by well-known prior studies~\cite{nguyen2018crosssim, nguyen2020automated}. This  dataset attempts to include repositories from ten different families
as listed in Table~\ref{tab:20_cluster}.
% such as PDF  processors, JSON parsers, Object Relational Mapping projects, and Spring  MVC  related  tools. 
Second,  our \dmal dataset includes 13 types of malware families listed in the same table.
% (Trojan, Spoof, Rootkit, Ransomware, Keylogger, Backdoor, Virus, Botnet, DDOS, Worm, Spyware, Spam, and Sniff). 
% Furthermore, our embedding seems to be able to place these repositories in appropriate clusters (Section~\ref{sec:application}), meaning that it manages to distinguish among them. 
In the future, we intend to collect more repositories in our dataset and include
more programming languages. The key bottleneck is the creation of groundtruth. 
% Finally, recall that we focus on java repositories here, although we intend to conduct similar experiments with other languages in the future.

% On the other hand, due to GitHub API limitation, its difficult and time consuming to retrieve a large number of repositories in GitHub. We argue that dataset is fairly representative as it consists repositories from a number of different categories.

{\bf f. Should we consider the popularity metrics?} So far, we did not consider the popularity metrics of the repositories, such as the number of stars, watches, and forks. 
While we intend to examine what  information we can extract from such metrics,
we argue that they will mostly help in finding the representative or influential repositories. 
Our preliminary analysis  suggests that
popularity does not provide information w.r.t. the type of the repository.
%  We believe these popularity metrics are misleading to consider in similarity measurement approaches.
 As a proof of concept, we can consider an initial and a forked repository: they are most likely nearly identical, but their popularity metrics can vary significantly. 
%  First, a copied repository will not have similar pattern of behaviour to a original repository at the same time. 
% \miii{If you Check 5 popular and 5 non-popular by hand, we can also add a comment -- SUPER LOW priority}
%-------------------
\miii{Possibly delete the last or make ti more appealing...if we have the space.}

\section{Related Works}
\label{sec:related_works}
Studying the similarity among software repositories has gained significant attention in the last few years. Most studies differ from our approach in that: (a) they do not incorporate all types of data present in a repository, (b) they do not present a feature vector keeping the semantic meaning of the metadata, source code, and structure of a repository, and (c) their approaches are not suitable for other ML classification tasks such as repository family classification, malware and benign repository classification, etc. We discuss the related works briefly below.

{\bf a. Software similarity computation: }
The prior studies in software similarity computation can be classified mainly into three groups based on the data they use: (a) high level meta data~\cite{thung2012detecting, thung2013automated, chen2015simapp, zhang2017detecting}, (b) low level source code~\cite{kawaguchi2006mudablue, mcmillan2012detecting}, and (c) the combination of both high  and low level data~\cite{nguyen2018crosssim, nguyen2020automated}. 

In an earlier study~\cite{thung2012detecting}, authors utilize repository tags to compute the similarity among repositories written in different languages. Capturing the weights of tags present in a repository, they create the feature vector and apply cosine similarity to compute the similarity. Later,  \cite{thung2013automated} proposes a library recommendation method, LibRec, using association rule mining and collaborative filtering techniques. It searches for the similar repositories to recommend related libraries for developers. Another effort ~\cite{chen2015simapp} proposes SimApp to identify mobile applications with similar semantic requirements. A recent approach,  RepoPal~\cite{zhang2017detecting}, utilizes readme file, and stars property of \github repositories to compute the similarity between two repositories.

On the other hand, MUDABLUE~\cite{kawaguchi2006mudablue} is the first automatic approach to categorize the software repositories using Latent Semantic Analysis (LSA) on source code. Considering the source code as plain text, they create a identifiers-software matrix and apply LSA on it to compute the similarity. Later, another study~\cite{tian2009using} categorizes the software repositories applying Latent Dirichlet Allocation (LDA) on the source codes. A recent study named CLAN~\cite{mcmillan2012detecting} computes the similarity between repositories by representing the source code files as a term-document matrix (TDM) where every class represents a row and the repositories are the columns. 

Finally, a very recent study~\cite{nguyen2018crosssim, nguyen2020automated} proposes CrossSim, a graph based similarity computation approach using both high level star property and API call references in source code files in a repository. Utilizing the mutual relationship, they represent a set of repositories as a graph and compute the similar repositories of a given repository from the graph. However, their work is limited by the external library call which may fool as the similarity will largely depends on it. Another study~\cite{capiluppi2020detecting} has confirmed that CrossSim may identify dissimilarity based on external API usage while internally implementing similar functionalities.  

{\bf b. Embedding approaches: }
A recent advancement in Natural Language Processing (NLP) has opened a whole new way of feature representation, a neural network based feature learning approach for discrete objects. First, introduction of word2vec~\cite{mikolov2013distributed}, a continuous vector representation of words from very large corpus, has paved the way. Later, another study named doc2vec~\cite{le2014distributed} introduces a distributed representation of variable length paragraph or documents. More recently, the embedding concept is being shared in other domains and has gained enormous success in effective feature representation such as graph embedding~\cite{grover2016node2vec, narayanan2017graph2vec}, topic embedding~\cite{niu2015topic2vec}, tweet embedding~\cite{dhingra2016tweet2vec}, and code embedding~\cite{alon2019code2vec,hoang2020cc2vec,kang2019assessing,theeten2019import2vec}.

%graph2vec narayanan2017graph2vec 

%topic2vec niu2015topic2vec

%tweet2vec dhingra2016tweet2vec

%subgraph2vec narayanan2016subgraph2vec

%cc2vec hoang2020cc2vec

\section{Conclusions}
\label{sec:concl}

%%%%%%%%%%%%%%%%%%%%%%%%%%%%%%%%%%%%%%%
% AUTHOR: Christos Faloutsos
% INSTITUTION: CMU
% DATE: April 2019
% GOAL: to streamline the paper presentations
%%%%%%%%%%%%%%%%%%%%%%%%%%%%%%%%%%%%%%%

We present \method, an approach to represent a repository in an embedding vector utilizing data from three types of information sources: (a) metadata, (b) repository structure, and (c) source code available in a repository. The main idea is to aggregate the embedding representations from these three types of information. 

%The advantages of the method are 
Our work can be summarized in the following points:
\ben
\item {\bf A highly effective embedding:} \method is a comprehensive embedding approach, 
which enables us to determine similar repositories with 93\% precision.
\item {\bf Improving the state of the art:} Our approach
 outperforms the best known method, CrossSim, by a margin of  15\% in terms of precision.  Also, it  finds nearly  twice  as  many  Strongly  Similar  repositories  and  30\%  less False Positives.
\item {\bf Facilitating the identification of malware}: Our approach can classify the malware and benign repositories with 98\% precision outperforming previous studies.
\item {\bf Enabling meaningful clustering}: Our approach identifies a tree hierarchy of repositories that aligns well with their purpose and lineage.
\een
% \miii{If you want to focus on "advantages", you need to separate advantages/novelty of the method and applications}

In the future, 
we first plan to extend the work with a larger dataset and a more extensive ground truth dataset.
In fact, we would like to help develop a community-wide benchmark that will facilitate further research.
Second, we would like to extend our work to other programming languages, which hinges mostly on developing a code2vec capability for other languages.
It would be interesting to see if different languages lend themselves to embedding representations the same way we are able to do here with Java.

Finally, we intend to open-source our code and datasets to maximize the impact of our work and facilitate follow up research.

\bibliographystyle{IEEEtran}
%\bibliography{BIB/ref}
\bibliography{000paper}

% \reminder{
% \newpage
% \appendix
% \input{070appendix}
% \newpage
% \input{080listOfToDo}
% }

\end{document}